\documentclass[a4paper,11pt]{article}
\usepackage{epsfig}
\usepackage{graphicx}
\usepackage{cite}

\newcommand{\be}{\begin{equation}}
\newcommand{\ee}{\end{equation}}
\newcommand{\bea}{\begin{eqnarray}}
\newcommand{\eea}{\end{eqnarray}}
\newcommand{\beq}{\begin{equation}}
\newcommand{\eeq}{\end{equation}}

\begin{document}

\title{Characterization of mammographic masses using a gradient-based segmentation algorithm and a neural classifier}

\author{Pasquale Delogu$^a$, Maria Evelina Fantacci$^a$, Parnian Kasae$^b$ \\
and Alessandra Retico$^a$\footnote{Corresponding author: Alessandra Retico, Ph.D.; E-mail: alessandra.retico@df.unipi.it
}}
\date{}

\maketitle

\begin{enumerate}
\item[$^a$] Dipartimento di Fisica dell'Universit\`a di Pisa and INFN Sezione
di Pisa,\\ 
Largo Pontecorvo 3, 56127 Pisa, Italy.
\item[$^b$] The Abdus Salam International Center for Theoretical Physics,\\
Strada Costiera 11, P.O. Box 563, 34100 Trieste, Italy.
\end{enumerate}
\abstract{ Computerized methods have recently shown a great
potential in providing radiologists  with a second opinion about
the visual diagnosis of the  malignancy of mammographic masses.
The computer-aided diagnosis (CAD) system we developed for the
mass characterization is mainly based on a segmentation algorithm
and on the neural classification of several features computed on
the segmented mass. Mass-segmentation  plays a key role in most
computerized systems. Our technique is a gradient-based one,
showing the main characteristic  that no free parameters have been
evaluated on the dataset used in this analysis, thus it can
directly be applied to datasets acquired in different conditions
without any ad-hoc modification. 

A dataset of 226 masses (109 malignant and 117 benign) has been used in this
study. 
The segmentation algorithm 
works with a comparable efficiency both on malignant and benign
masses.
Sixteen features based on shape,
size and intensity of the 
segmented masses are extracted and
analyzed by a multi-layered perceptron neural network
trained with the error back-propagation algorithm.  
The capability of the system in discriminating malignant
from benign masses has been evaluated in terms of the
receiver-operating characteristic (ROC) analysis. 
A feature selection procedure has been carried out on the basis
of the feature 
discriminating power and of the linear correlations interplaying 
among them.
The comparison of the areas under 
the ROC curves obtained by varying the number of features to be classified 
 has shown that 12 selected features out of the 16 computed ones 
are powerful enough to achieve the best  classifier performances.
%
The radiologist 
assigned the segmented masses to three different categories: {\it  correctly-}, 
{\it acceptably-} and {\it non-acceptably-segmented} masses.
We initially    
estimated the area under ROC curve only on
the first category of segmented masses (the 88.5\% of the dataset), then extending the classification to the 
second subclass (reaching the 97.8\% of the dataset) and finally to the whole dataset, obtaining 
$A_z = 0.805\pm 0.030,  0.787\pm 0.024$ and 
 $0.780\pm 0.023$, respectively. 
%
\\

}

\section*{Introduction}

Breast cancer is still one of the most common forms of cancer
among women, despite earlier detection and more effective
treatments have contributed to a significant decrease in the
breast-cancer mortality during the last
decades~\cite{Greenlee,Levi,Smith,Landis}. Mammography is widely
recognized as the most reliable technique for early detection
of breast cancers~\cite{Zuckerman,Haus}. Once a mass is detected
on a mammogram, the radiologist recommends further investigations,
depending on the probability of malignancy he assigns to that
lesion. However, the characterization of masses from mammographic
images is a very difficult task and a high number of unnecessary
biopsies are actually performed in the routine clinical activity.
The rate of positive findings for cancers at biopsy ranges from
15\% to 30\%~\cite{Adler}, i.e. the specificity in differentiating
malignant from benign lesions on mammographic images is rather
low. As a breast biopsy is an invasive and expensive procedure,
methods to improve mammographic specificity without missing cancer
have to be developed. A higher predictive rate of the mammographic
examination can be achieved by combining the radiologist's
interpretation and the computer analysis. Computerized method have
recently shown a great potential in assisting radiologists in the
malignant or benign decision, by providing them with a second
opinion about the visual diagnosis of the
lesion~\cite{Huo1,Huo2,Sahiner0}.

The computer-aided diagnosis (CAD) system for characterizing
masses described in this paper is based on a three-stage
algorithm: first, a segmentation technique extracts the mass from
the image; then, several features based on size and shape of the
lesion are computed; finally,  a neural classifier merges the
features into a likelihood of
 malignancy for that lesion.
With respect to a number of CAD systems with a similar purpose and using a similar approach  already discussed in the literature,
the system we present shows the distinguishing characteristic
that a robust segmentation technique has been implemented:
it is based on a segmentation algorithm completely free
from any application-dependent parameter.

This paper is structured as follows: the methodology is presented in sec.~\ref{sec:method}, sec.~\ref{sec:dataset} describes the mammographic dataset available for our study and sec.~\ref{sec:results} reports on the analysis details and on the whole  system performances.

\section{Methodology}
\label{sec:method}

\subsection{Mass segmentation}

Mass segmentation is a quite difficult task because masses are
often varying in size, shape and density. Masses can exhibit a
very poor image contrast or can be highly connected to the
surrounding parenchymal tissue. Thus, it is hard in many cases to
distinguish the mass from the nonuniform normal breast tissue. Due
to the high variability in the appearance of masses, generalizing
a segmentation algorithm able to handle many different types of
masses is a nontrivial task and much efforts have already gone
through this issue~\cite{Wirth,Amini,Sahiner,Matthew,Timp}.

\begin{figure}
\begin{center}
\epsfxsize 13. true cm
\epsffile{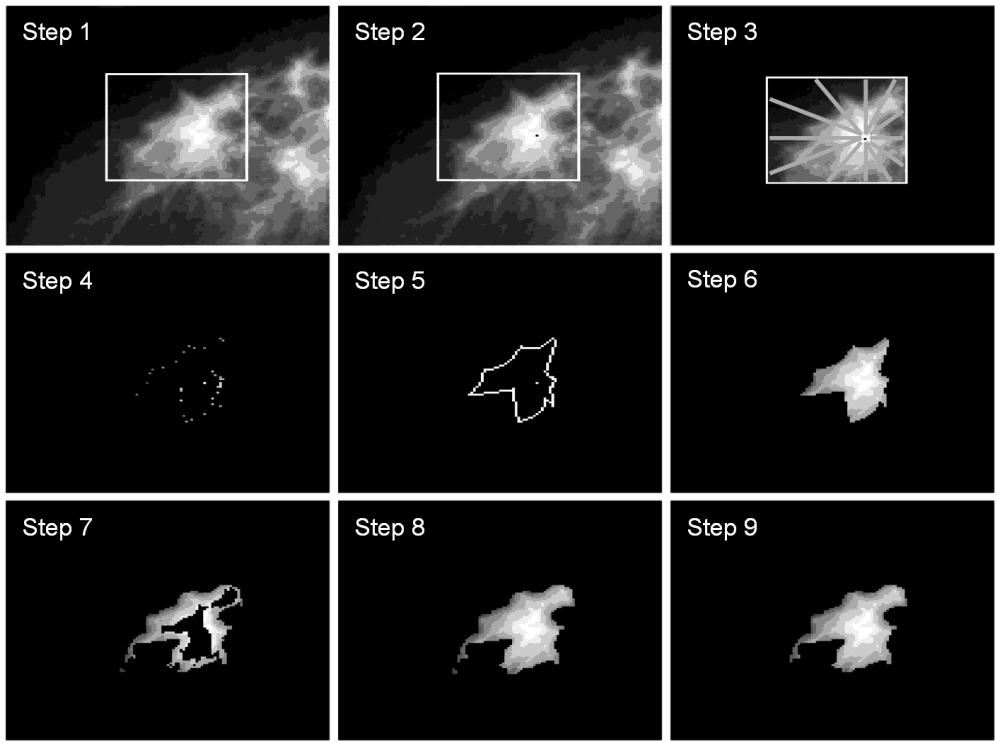}
\end{center}
\caption{\label{fig:segmentation} Mass segmentation procedure: for the explanation of the algorithms implemented in each single step see the text.
}
\end{figure}

The segmentation algorithm we developed is an extension and a
refinement of the strategy proposed in~\cite{Chen} for the
mass segmentation in the CAD analysis of breast tumors on
sonograms.
The procedure we propose is able to identify the mass shape within a  Region Of
Interest (ROI)  the radiologist interactively chooses  
on the mammogram.
Despite the radiologist is asked to select  the smallest 
region containing the mass, the ROIs usually contain the lesions as well as a 
considerable part of normal tissue.
Our segmentation method  aims at removing the non-tumor
regions around the tumor in a ROI by applying the following
processing steps (see fig.~\ref{fig:segmentation}).

\begin{description}
\item[Step 1.] Once the ROI containing the mass has been selected by the physician, 
a shrinking factor of 8 is applied on both rows and columns within the rectangular region both to reduce the high-frequency 
noise affecting the digitized images and to limit the segmentation computing time. 
\item[Step 2.] Assuming that masses become denser and denser as
going from the boundary to the center, we took the pixel with the
maximum-intensity value as the starting point for the segmentation
algorithm (seed point). Since the ROI can even contain pixels
belonging to the normal tissue with higher intensity values with
respect to the pixels representing the mass, the center of the ROI
is taken as the seed point if the distance between the
maximum-intensity pixel and the center of the ROI exceed the 75\%
of the ROI half diagonal\footnote{As the ROI is a user-drawn
rectangle containing the mass, the center of the ROI and the
center of the lesion are not expected to be very different.}.
\item[Step 3.] A number of radial lines are depicted from the
chosen seed point to the boundary of the ROI area. 
\item[Step 4.]
We scan the pixels along each radial line starting from the center
to the boundary of the ROI. We look for the pixel whose local
variance is maximal. The local variance is defined as the variance
in a predefined  $n \times n$  matrix with the currently processed
pixel in the center. The pixel with the maximal local variance is
considered to be most likely the boundary point between the mass
and the surrounding tissue: it is referred in what follows as
critical point. The smaller the size of the matrix we consider,
the more sensitive to the small variances or details. This can be
a helpful parameter for detecting the arms and branches of the
stellate masses. Both to enhance the sensitivity of the segmentation
procedure to small details of the lesions and to reduce the 
algorithm execution time, the smallest size of the pixel neighborhood  
has been chosen in the computation of the local variance, i.e. $n=3$.
\item[Step 5.] After scanning all radial lines and finding the
critical points corresponding to each one of them, the critical
points are linearly interpolated. 
\item[Step 6.] The region inside
the coarse boundary so far identified is filled. We use the pixels
of this region as seed points for the further steps of the
segmentation algorithm, whose aim is to lead to a more detailed
and more accurate identification of the shape of the lesion.
\item[Step 7.] The steps 3 and 4 are iterated for each seed point
identified in the step 6. What we obtain is a set of points
detected from different angles to be most probably located on the
boundary. In order to select the right thin boundary out of this
set, we first tried to assign a vote to the pixel  detected at
each time as a critical point so that the higher the credit of the
critical point, the higher the probability of being appropriate to
represent the real boundary of the mass. We found out that
selecting appropriate criteria for assigning and thresholding the
votes was crucial for the mass border identification and led to
disconnectivity. Therefore, in order to avoid the unnecessary
presence of free application-dependent parameters in this
procedure we decided to accept all identified points. In this way
we will end up with a thick and more connected border. 
\item[Step 8.] To complete the identification of the mass, the area inside
the border has to be filled. Since we may have self-intersecting
region it is more convenient to fill the background. To prevent
entering the mass from possible disconnectivity in the border we
use a cross-like mask for filling background. Subtracting
filled background from ROI will give the mass. 
\item[Step 9.] A final filtering is performed in order to remove some possibly
present non-connected objects.
\end{description}

\subsection{Feature extraction}
\label{sec:feature_extraction}

Once the masses have been segmented out from the surrounding normal
tissue, a set of 
morphological and textural
features are computed in order to
allow a decision-making system to distinguish benign from malignant
lesions.
The likelihood of malignancy for a mass can in fact be 
estimated
on the basis of its morphological  and textural appearance, which is usually 
described in terms of the mass  size, shape,
margin characteristics and x-ray attenuation 
(radio-density)~\cite{Wang,Kinnard,Christoyianni,Qian,Hadjiiski,Sahiner2,Mudigonda,Huo}.

Despite mass size alone does not predict malignancy, the
size of a malignant mass is indicative of its progression. Therefore
features like \textit{area} and \textit{perimeter} are usually included
in the set of features to be computed.

The mass shape can be round, oval, lobular or irregular. 
Features like: \textit{circularity, convexity, maximum axis, minimum axis}
can be useful in mass malignancy definition since most benign masses appear circular 
and convex whereas malignant cases have irregular non-convex shapes. 

The study of the mass margin characteristics is probably the most important 
in determining whether the mass is likely to be benign or
malignant. There are five type of mass margins as defined by 
BI-RADS\textregistered~\cite{Birads}:
circumscribed, obscured, micro-lobulated, ill-defined, and
spiculated. Circumscribed margins are well defined and sharply
demarcated with an abrupt transition between the lesion and the
surrounding tissue. Microlobulated margins have small undulating
circles along the edge of the mass. Obscured margins are hidden by
superimposed or adjacent normal tissue. Ill-defined margins are
poorly defined and scattered. Spiculated margins are marked by
radiating thin lines. 
 The features which can estimate the level of
spiculations and evaluate the softness or roughness of the margin
are: \textit{mean and  standard deviation of normalized radial length, 
 radial length entropy, zero crossing,
mean and standard deviation of the variation ratio.}

The X-ray attenuation is a description of the density of the
mass. Breast cancer often appears denser, i.e. whiter, than the
surrounding normal breast parenchyma. The intensity
and its variation inside the mass can be measured by features like: 
\textit{mean, standard deviation, kurtosis and skewness of the 
mass intensity.}

Supported by the existing correlations between the morphological and
textural features of a mass and its likelihood of malignancy, 
we computed the already mentioned 16  features on segmented masses 
according to the  following formulas.
 

\begin{enumerate}
\item
{\bf Mass Area}: it is given by the number of pixels inside the boundary of the mass.
\item
{\bf Mass Perimeter}: it is measured by summing up the
number of pixels on the boundary of the mass.
\item
{\bf Circularity}:
\beq
C =\frac{4\pi A}{P^2},
\eeq
where $P $ is the perimeter and $A$ is the area of the mass.
The circularity $C$ is
calculated in such a way that for a mass with circular shape $C=1$.
Changing the shape to oval or irregular this number decreases.
\item
{\bf Mean of the Normalized Radial Length}:
\beq
d_{\rm avg} =\frac{1}{P} \sum_{i=1}^P d(i),
\eeq
where $d(i)$ is the Euclidean distance from the center of mass of the segmented
lesion to the $i^{\rm th}$ pixel on the mass boundary and normalized with respect to the maximum distance found for that mass. $P$ is the mass perimeter.
\item
{\bf Standard Deviation of Normalized Radial Length}:
\beq
\sigma = \sqrt{\frac{1}{P} \sum_{i=1}^P \left( d(i)-d_{\rm avg}\right)^2 }.
\eeq
It is a good measure for irregularity. The more spiculations and
irregularities are present, the higher is the standard deviation of the
radial length.
\item
{\bf Radial Length Entropy}: it is a probabilistic measure computed from the
histogram of the normalized radial length as follows:
\beq
E = - \sum_{k=1}^{N_{\rm bins}} P_k \log{P_k}.
\eeq
The parameter  $P_k$ is the probability that the normalized radial length   is
between $d(i)$ and $d(i) + 1/N_{\rm bins}$, where $N_{\rm bins}$ is the  number of bins the normalized histogram, ranging in the [0,1] interval, has been divided in ($N_{\rm bins}=5$ in our analysis).
\item
{\bf Zero Crossing}: it is a count of the number of times
the radial distance plot crosses the average radial distance. It is an
indicator for the degree of spiculation of the mass.
\item
{\bf Maximum Axis}: it consists is the largest distance
connecting one point along the mass boundary to another point on the
mass boundary going through the center of mass of the lesion.
\item
{\bf Minimum Axis}: it is the shortest distance
connecting one point along the mass boundary to another point on the
mass boundary going through the  center of mass of the lesion.
\item
{\bf Mean of the Variation Ratio}: first we find the variation of all radial
length from their mean value, then we determine the maximum variation
magnitude $var_{\rm max}$ of radial length. Only those variations
having a magnitude greater than $var_{\rm max}/2 $ are considered as
dominant variations.
The variation ratio mean is computed as
the average of those dominant variations.
\item
{\bf Standard Deviation of the Variation Ratio}: it is calculated as the
 standard deviation of the dominant variations with respect to the variation
ratio mean. It indicates the sharpness of the variations and
can be a good indicator for spiculation.
\item
{\bf Convexity}: it is the ratio between the mass area and the area
of the smallest convex containing the mass.
If the mass has a regular shape, which
means it is convex, this number is one, otherwise it
will decrease.
\item
{\bf Mean of the Mass Intensity}: it is the mean value of the grey-level intensity values of the pixels inside the mass boundaries.
\item
{\bf Standard Deviation of  the Mass Intensity}:
it is a measure of the smoothness of the grey-level intensity  values of the pixels  inside the mass boundaries.
\item
{\bf Kurtosis of the Mass Intensity}: it
is a measure of how outlier-prone is a distribution. It is defined as follows:
\beq
kurtosis = \frac{E\left( g(i,j) - \mu\right)^4}{\sigma^4},
\eeq
where $g(i,j)$ is the grey level at location $(i, j)$, $\mu$ and $\sigma$ are the average
intensity and standard deviation inside the segmented mass, respectively.
We can investigate how far is the
intensity distribution of the mass from a normal distribution.
The {\it kurtosis} of the
normal distribution is 3. Distributions that are more outlier-prone than
the normal distribution have a {\it kurtosis} greater than 3; distributions that are less outlier-prone have a {\it kurtosis} lower than 3.
\item
{\bf Skewness of the Mass Intensity}: it is a measure of the
asymmetry of the data around the sample mean. It is given by:
\beq
skewness = \frac{E\left( g(i,j) - \mu\right)^3}{\sigma^3}.
\eeq
If the {\it skewness} is negative,
the data are spread out more to the left of the mean than to
the right. If it is positive, the data are spread out more to
the right. The  {\it skewness} of a normal distribution or a perfectly
symmetric distribution is zero.
\end{enumerate}

\subsection{Classification}

Once the features are extracted from the segmented masses, one faces with 
the choice of an appropriate classification method.
Lots of different approaches implemented to this purpose have been widely discussed in literature,
such as the \textit{Minimum Distance Classifier}~\cite{Lambrou,Bruce}, the
\textit{K-Nearest Neighbor Distance Classifier}~\cite{Lambrou}, the
\textit{Linear Discriminant Method (LDA)}~\cite{Wei,Bruce,Sahiner,Hadjiski2}, the
 \textit{K-mean clustering}~\cite{Lambrou}, the \textit{Binary Classification Tree}~\cite{Sheng}
 and \textit{Artificial Neural Network (ANN)}~\cite{Wang,Arbach,Bovis,Lauria,Lim}.
Neural networks are widely used because of their capability of simultaneously
processing large amounts of information, for their
 ability in analyzing and classifying patterns even when presented with 
noisy or partial information and to adapt their behavior to the nature
of the training data. 

Relying on the 
advantages of a neural approach,  we implemented a supervised neural classifier 
in our CAD scheme.
A standard three-layer feed-forward neural network~\cite{Haykin}
has been chosen to this purpose.
The general architecture of this ANN
consists in $n$ input, $h$ hidden and two output neurons,
and the supervised training phase
is based on the back-propagation algorithm.
We used the sigmoid activation function for both the hidden
layer and the output layer, and the on-line learning method allowing weights
to be updated after the presentation of each pattern.
Updating was synchronous, therefore all
nodes were updated at the same time.
The supervised learning is performed by showing the network a set
of training vectors constituted by the input pattern and the corresponding
target response.
In particular, the target $[1,0]$ has been added to the vectors
of features extracted from malignant masses, whereas
the target $[0,1]$ to those derived from benign masses.
Let us assume $[y_1, y_2 ]$ to be the output of the network
for a given input:
the corresponding lesion is classified as malignant if
$y_1>y_2$, otherwise it is assumed to be benign.
Whereas the number $n$ of units in the input layer is  {\it a priori} fixed by 
the choice of the number of features to by classified,
the number $h$ of hidden neurons has to be
experimentally determined on the basis of the training dataset.

The performances of the training algorithm
were evaluated according to the 5$\times$2 cross validation
method~\cite{Dietterich}. It is the recommended test to be performed on
algorithms that can be executed 10 times because it can provide a reliable
estimate of the variation of the algorithm performances
due to the choice of the training set.
This method consists in performing 5 replications of the 2-fold cross
validation method~\cite{Stone}.
At each replication, the available data are randomly partitioned
into 2 sets ($A_i$ and $B_i$ for $i=1,\dots5$) with an almost equal
number of entries.
The learning algorithm is trained on each set and
tested on the other one.
The performances the system achieves in the classification phase are given
in terms of the sensitivity and specificity values.
The sensitivity is defined as
the true positive fraction (fraction of malignant masses correctly classified by the system), whereas the specificity is referred to the true negative
fraction (fraction of benign masses correctly classified by the system).

The performances of the neural classifier were also evaluated in terms of a
Receiver Operating Characteristic (ROC) analysis~\cite{Metz}.
In order to show the trade off
between the sensitivity and the specificity, a
ROC curve is obtained by plotting the
true positive fraction
versus the false positive fraction
of the cases (1 - specificity), computed
while the decision threshold of the classifier is varied.
Each decision threshold results in a corresponding
operating point on the curve, which usually
goes through the points $(0,0)$, where the classifier
detects no positives, and $(1,1)$, where every pattern is classified as
positive.

\section{Image data set}
\label{sec:dataset}

The image data set used for this study has been extracted
from a large database of mammograms
collected in the framework of a Collaboration between physicists from
several Italian Universities and INFN (Istituto Nazionale di Fisica Nucleare)
Sections, and radiologists from several Italian
Hospitals~\cite{Bottigli,magic5}.
The mammograms come both from screening and from the routine work carried out in the participating Hospitals.
The $18 \times 24$ cm$^2$ mammographic films were
digitized by a CCD linear scanner
(Linotype Hell, Saphir X-ray).
The digitized images are characterized by
a  $85 \mu$m pixel pitch and a 12-bit resolution, thus allowing up to
4096 grey levels.
The pathological images are fully characterized by a consistent
description, including the radiological diagnosis, the
histological data and  the coordinates of the center
and the approximate
radius (in pixel units) of a circle enclosing the masses ({\it truth circle}).
Mammograms with no sign of pathology are stored as normal images only
after a follow up of at least three years.

\begin{figure}
\begin{center}
\epsfxsize 10.  true cm
\epsffile{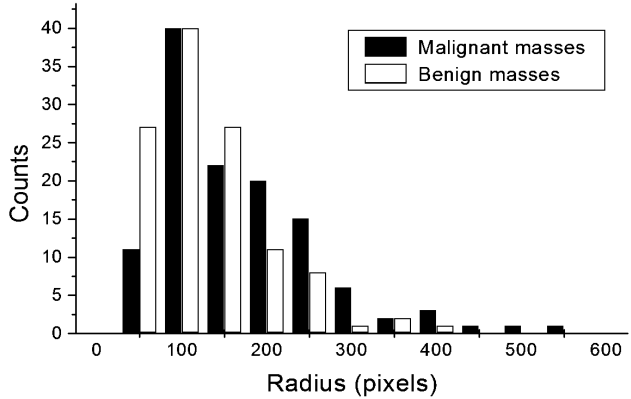}
\end{center}
\caption{\label{fig:histoMalBen}
Distribution of the mass sizes: the radius of the {\it truth circles} annotated by experienced radiologists are shown for malignant and benign masses.
}
\end{figure}

A set of 226 masses were used in this study: 109 malignant and 117
benign masses were extracted from single-view cranio-caudal or
lateral  mammograms.
The distribution of the mass sizes  can be observed in fig.~\ref{fig:histoMalBen}, where the histograms of the radius in pixels of the {\it truth circles} indicating the pathological regions, as annotated by experienced radiologists, are shown for the malignant and benign cases.
The diameters of the {\it truth circles} in real units are in the range 6.6--94.4~mm. It is worth noting that the size of the {\it truth circle} usually overestimates the real size of the mass.
The dataset we analyzed can be considered as representative of the patient population that is sent for biopsy under the current clinical criteria.

\section{Results}
\label{sec:results}

\subsection{Segmentation Result}

The  segmentation algorithm efficiency was directly
evaluated with the assistance of an experienced radiologist who, 
after selecting the ROI to be analyzed~\footnote{The current  CAD GUI allows 
the radiologist to select a maximum area of 1000$\times$1000 pixels 
on the digitized mammogram. Despite this technical restriction could easily be removed, it does not affect the current analysis as the maximum actual size of the larger mass in our database is less than 1000 pixels. 
The disagreement of this statement with the histogram reported in 
fig.~\ref{fig:histoMalBen} is only apparent because the  radiologist's {\it truth circles} are usually very conservative; a factor of $0.8\pm 0.3$ actually occurs between  half of the maximum mass axis (computed on the  segmented masses and confirmed by an experienced radiologist) and the radiologist's annotated
mass radius for the  {\it truth circles}.}, 
assigned
the  masses automatically segmented by the system to one out of the 
following three categories:  {\it correctly-segmented}, {\it acceptably-segmented} or {\it non-correctly-segmented} masses.

\begin{figure}
\begin{center}
\epsfxsize 13. true cm
\epsffile{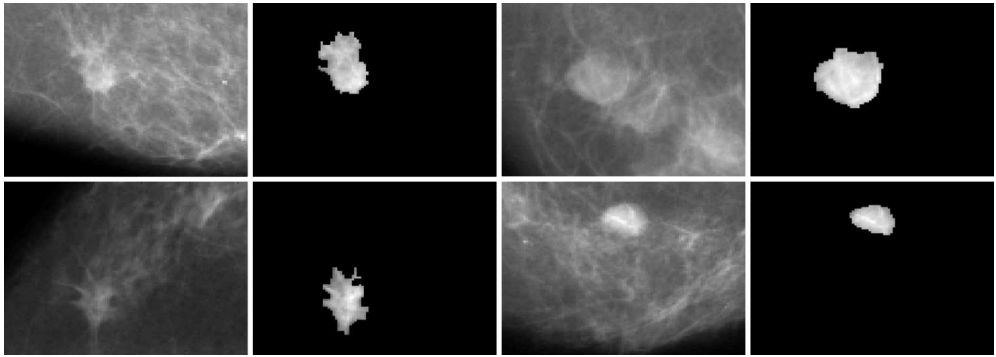}
\end{center}
\caption{\label{fig:segm_examples}
Examples of {\it correctly-segmented} masses: two malignant masses on the left and two benign masses on the right.
}
\end{figure}

\begin{figure}
\begin{center}
\epsfxsize 13. true cm
\epsffile{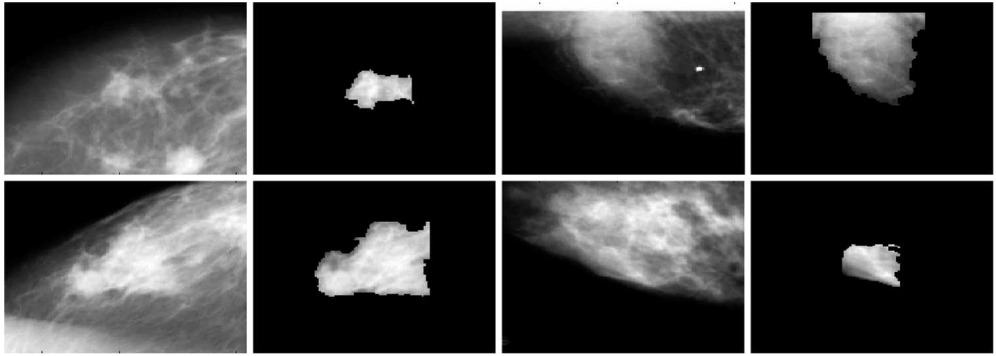} 
\end{center}
\caption{\label{fig:segm_examples_quasiOK}
Examples of  {\it acceptably-segmented} masses: two malignant masses on the left  and two benign masses on the right.
}
\end{figure}

The radiologist has been asked to classify as {\it correctly segmented} only those  masses
whose identified boundary was sufficiently close to that she would have drawn
by hand on the image. 
Despite the borders of the masses are usually not very sharp in
mammographic images, the segmentation procedure we propose leads in most cases
to a quite accurate identification of the mass shapes (see fig.~\ref{fig:segm_examples}). 
In fact, 200 masses (95 malignant and 105 benign) out of the dataset of 226 cases, were correctly segmented leading to an efficiency of correct segmentation   $ \epsilon_{\rm CS} = 88.5\%$. 
%
%
Out of the 26 remaining cases, 21 masses were assigned to the category of 
 {\it acceptably-segmented} masses, whereas 5 masses where definitely rejected by the radiologist as  {\it non-correctly-segmented} cases.
Some examples of  reasonably segmented masses are shown in fig.~\ref{fig:segm_examples_quasiOK}. 
It can be noticed that these masses are usually characterized by an obscured margin or they are not fully visible in the available mammographic 
field of view. 
In such cases the interactive selection of an appropriate ROI becomes particularly difficult and user-dependent, ending up with a 
non-satisfactory identification of at least a  portion of the mass margin.
In case of {\it non-correctly-segmented} masses a too large portion of the mass margin is not correctly  identified.
If the {\it acceptably-segmented} masses are added to the set of {\it correctly-segmented} masses the fraction of the dataset of masses that will be analyzed and classified by the CAD system will reach the value $ \epsilon_{\rm CS+AS} = 97.8\%$.


\subsection{Analysis of the extracted features}

Despite each of the 16 features we computed on the segmented masses 
can potentially enlighten a different characteristic of a mass
and contribute to a good classification result,
we  performed some tests to evaluate the discriminatory 
power of each feature 
and the degree of linear correlation among the different features.
We restricted this analysis  to the feature extracted from the
200 {\it correctly-segmented} masses.

\begin{figure}
\begin{center}
\epsfxsize 6.5  true cm
\epsffile{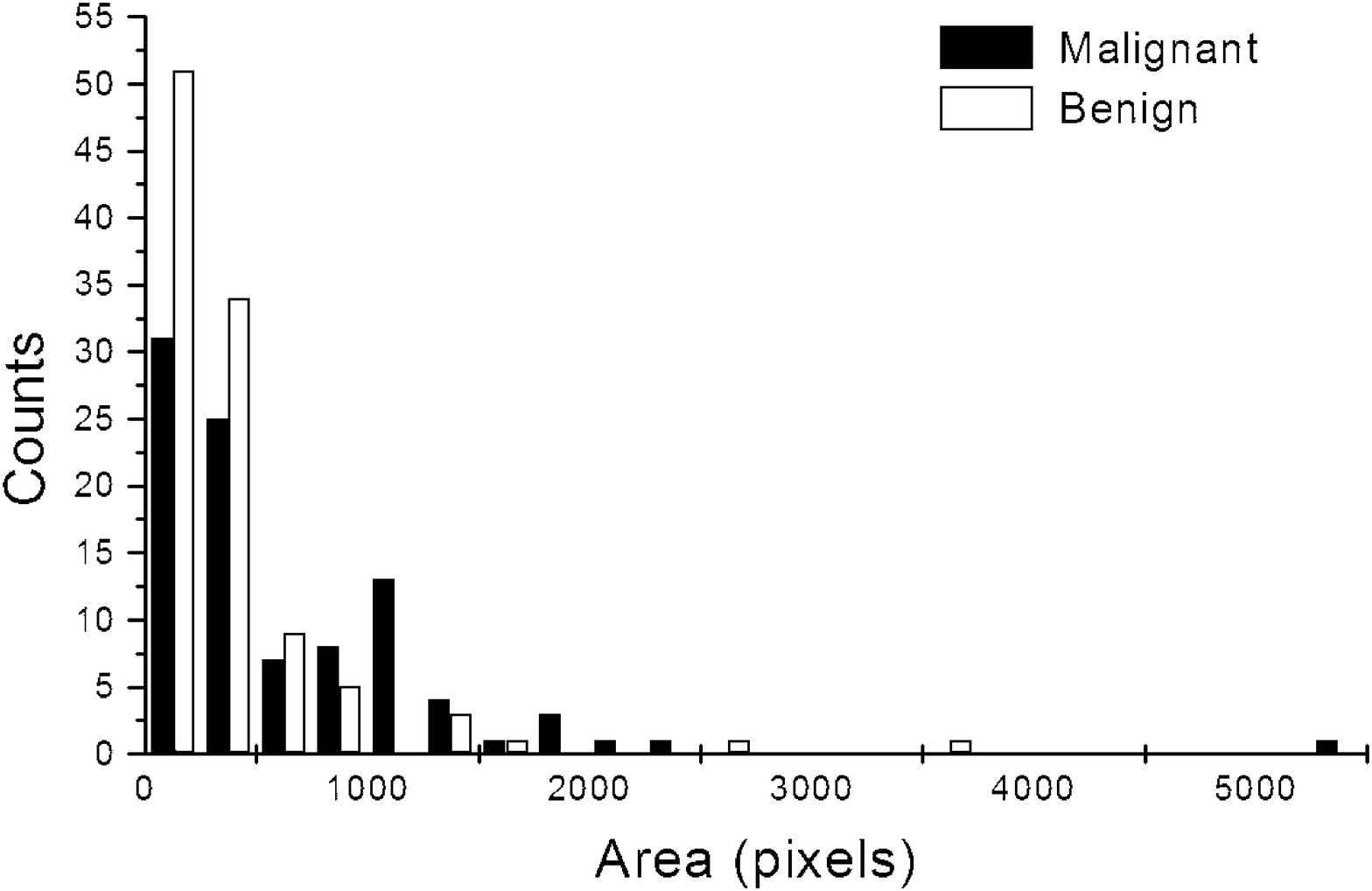}
\epsfxsize 6.5  true cm
\epsffile{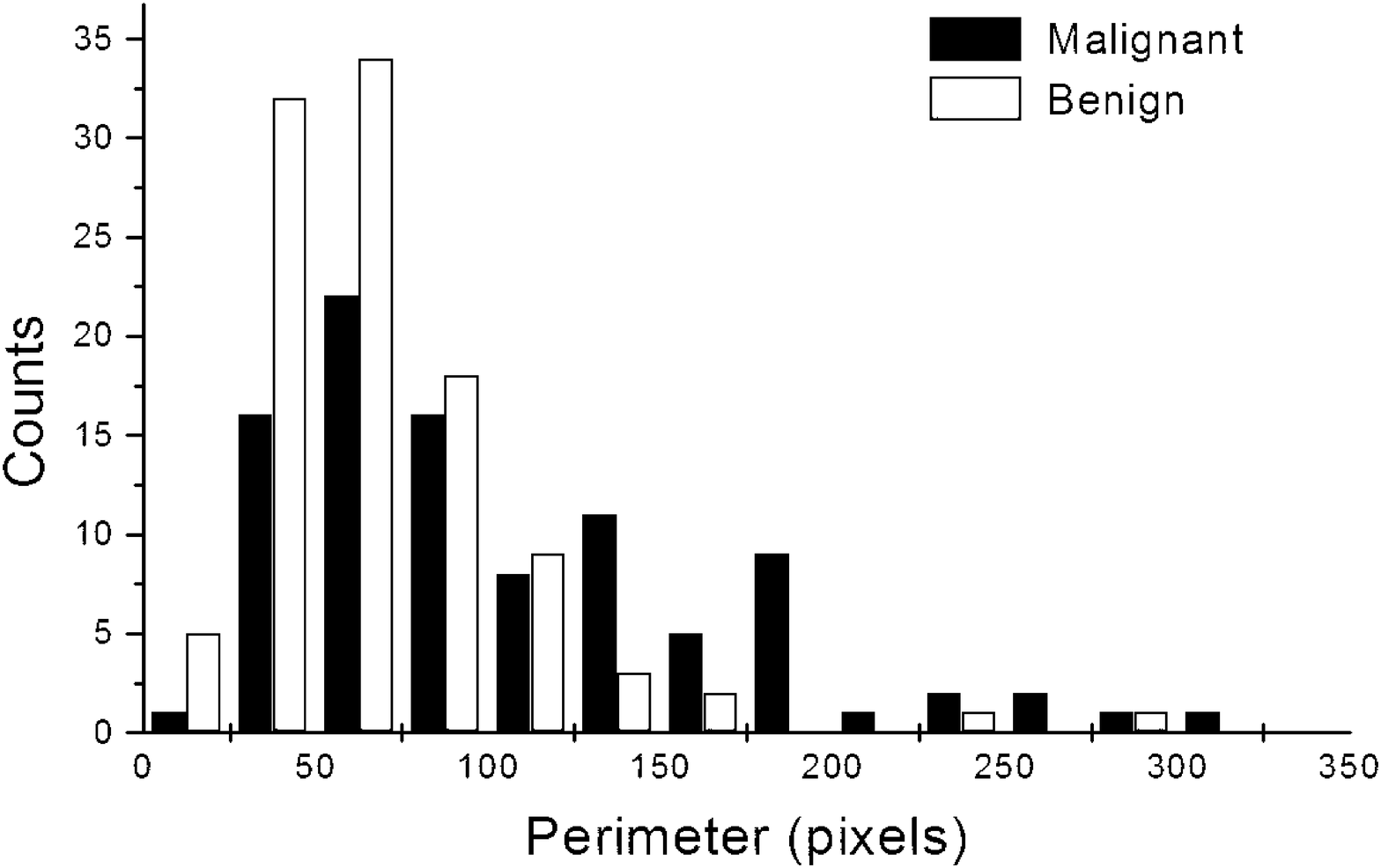}\\
\epsfxsize 6.5  true cm
\epsffile{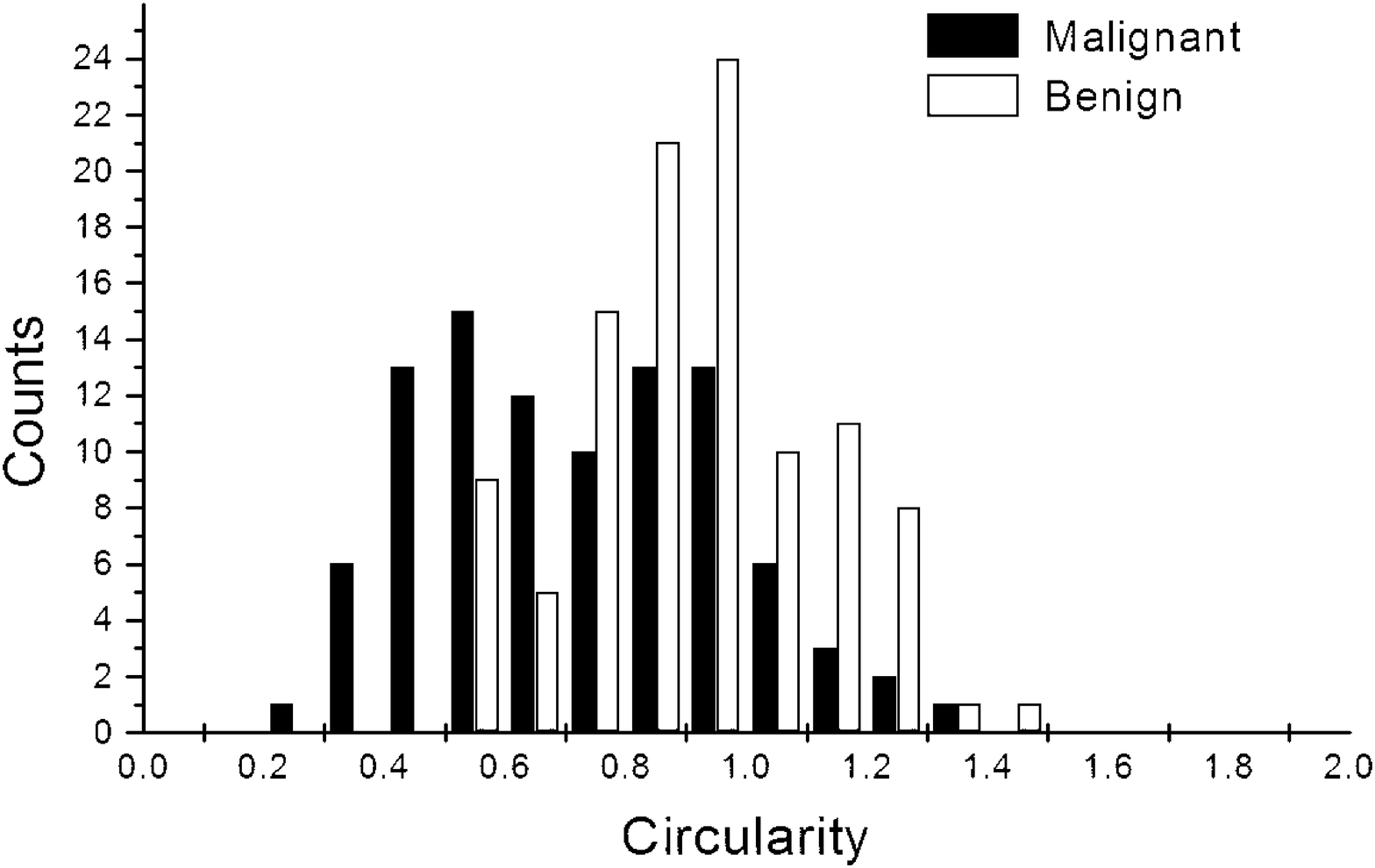}
\epsfxsize 6.5  true cm
\epsffile{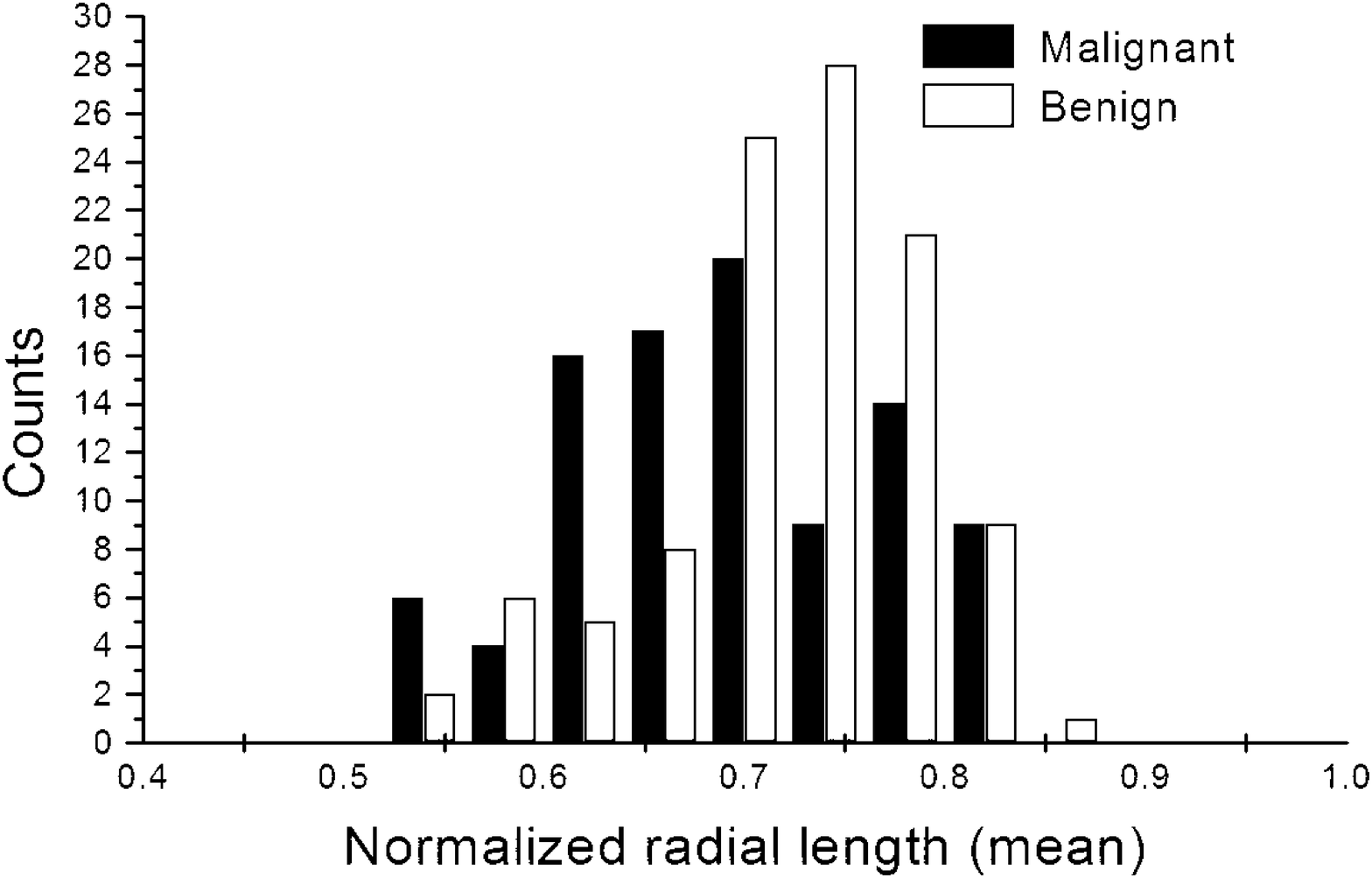}\\
\epsfxsize 6.5  true cm
\epsffile{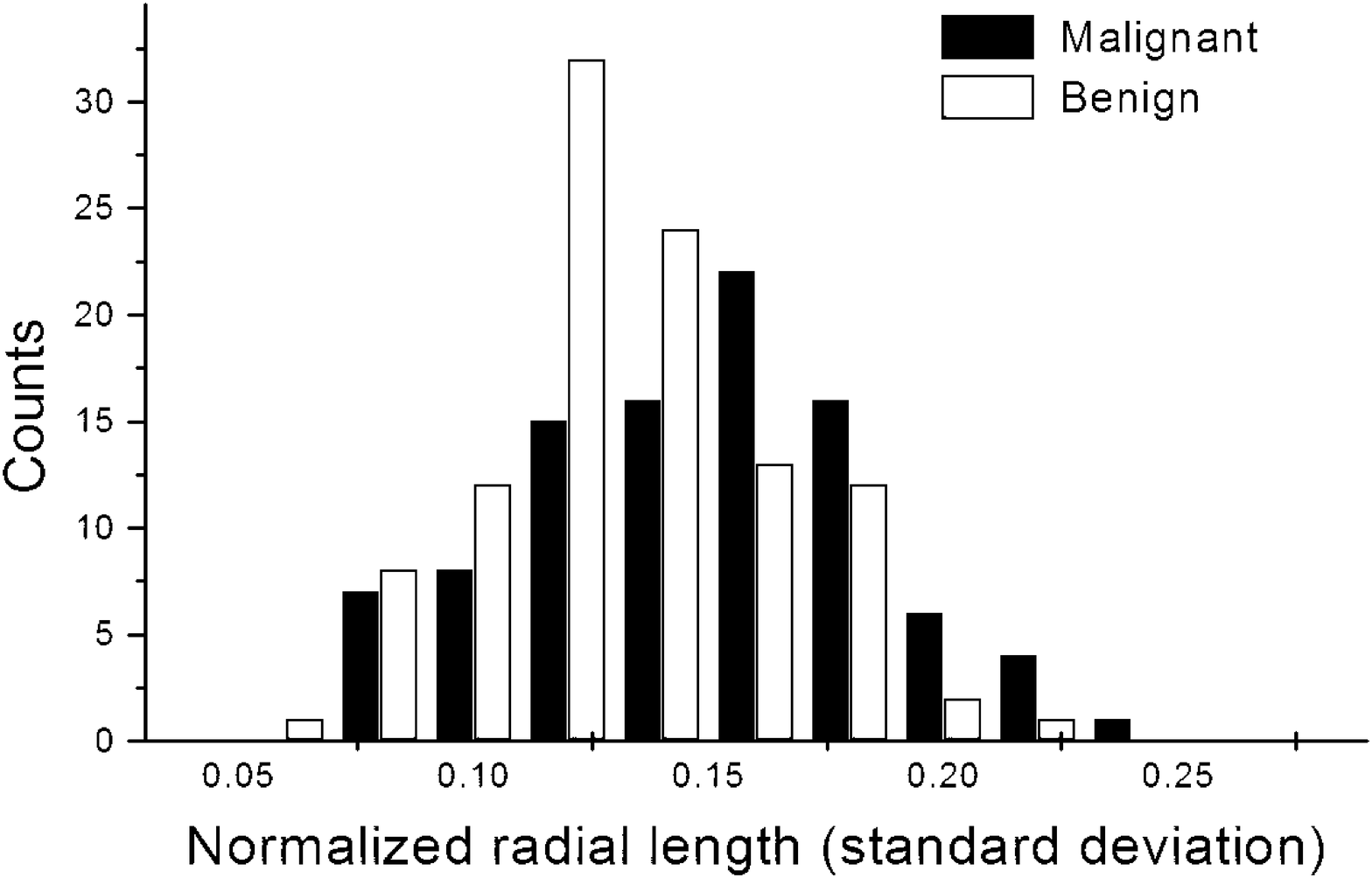}
\epsfxsize 6.5  true cm
\epsffile{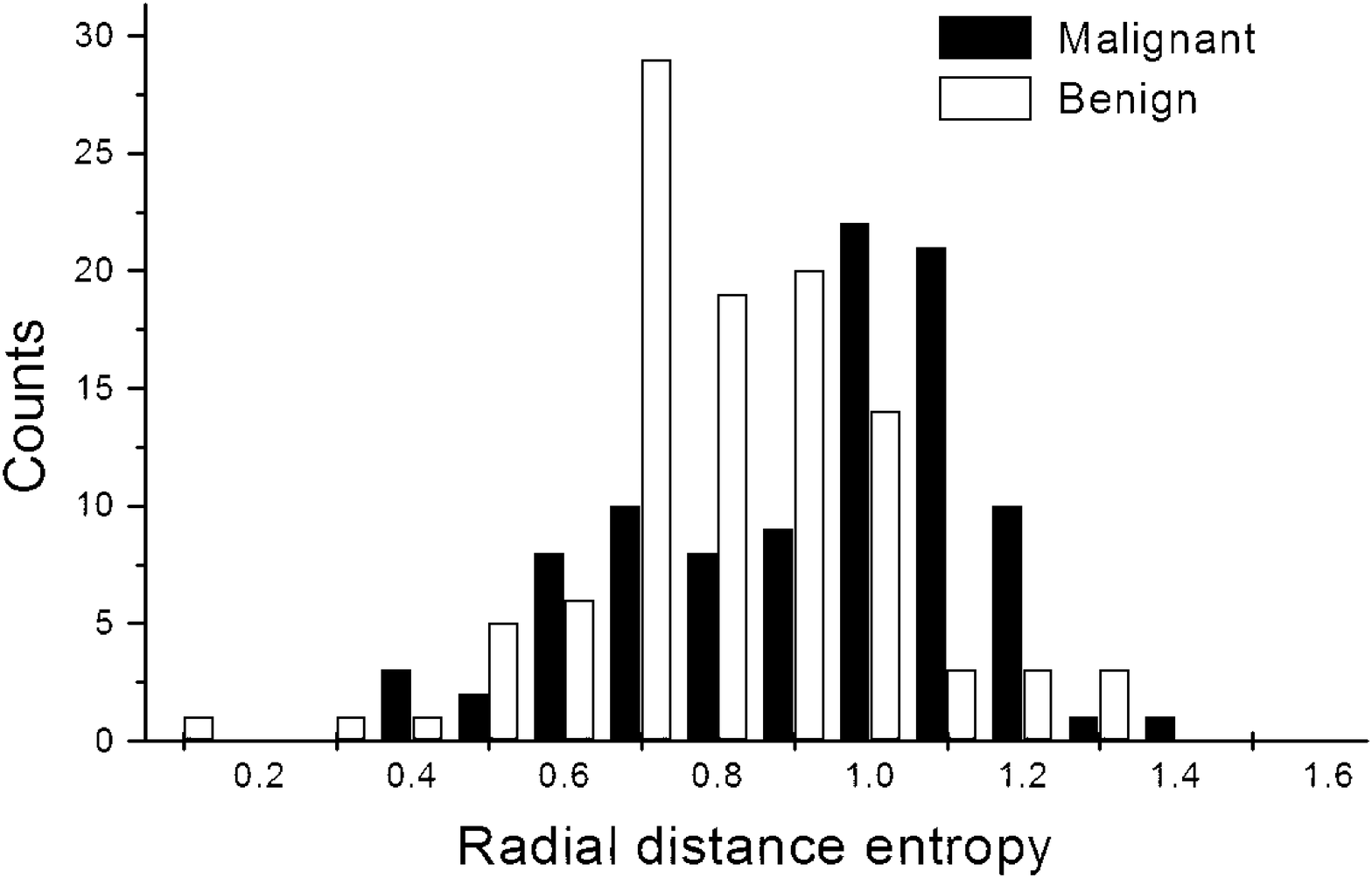}\\
\epsfxsize 6.5  true cm
\epsffile{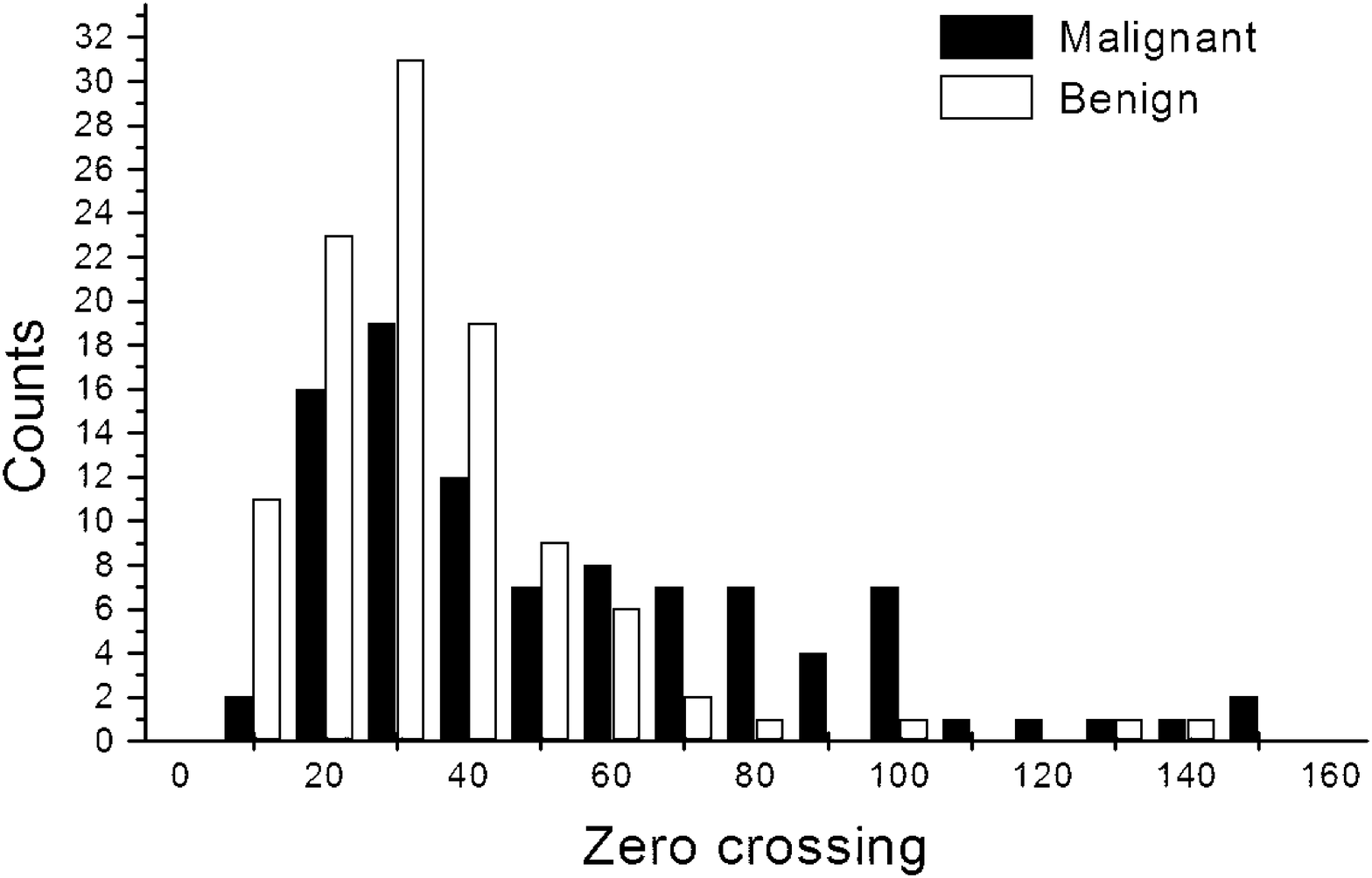}
\epsfxsize 6.5  true cm
\epsffile{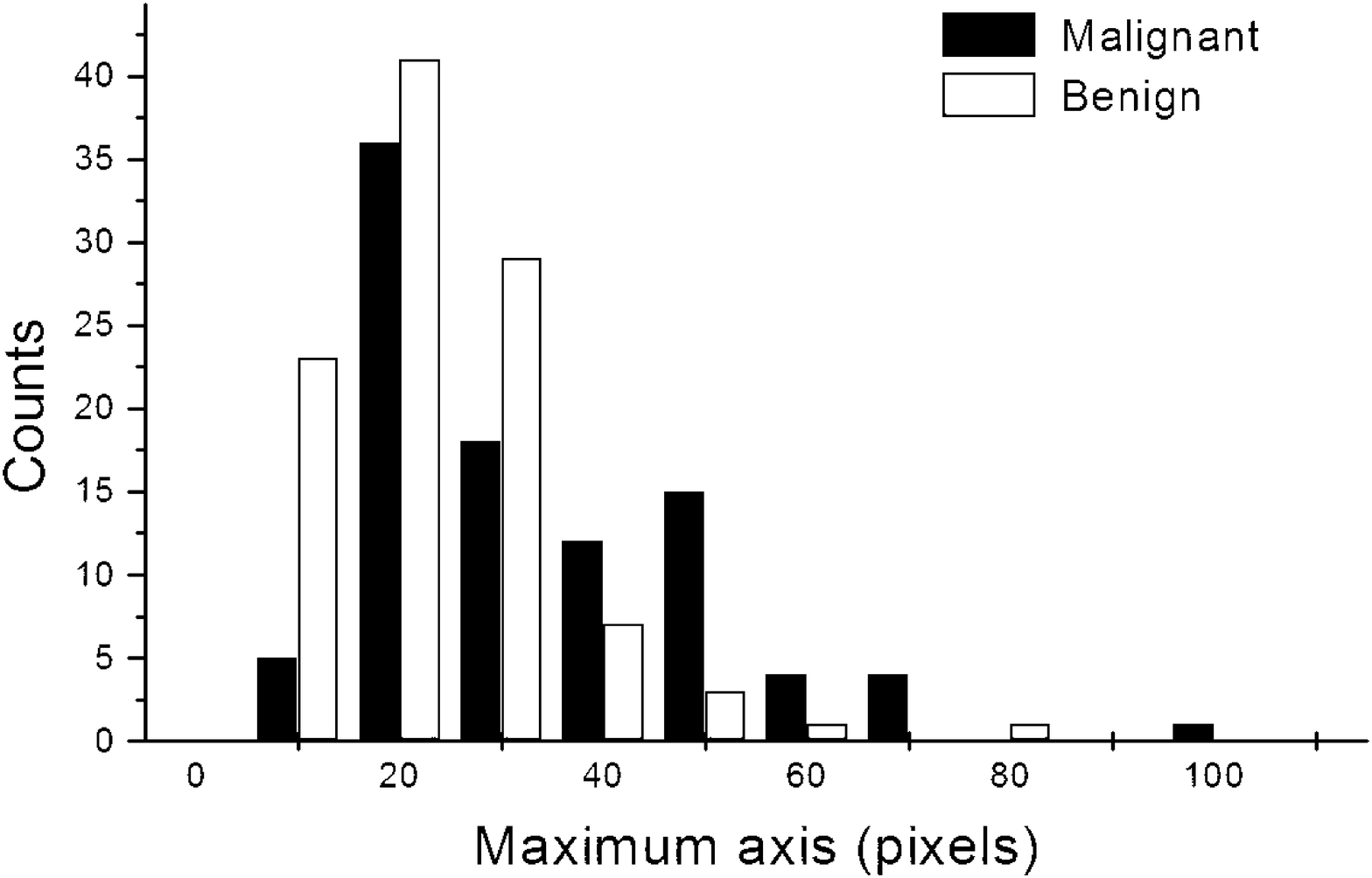}
\end{center}
\caption{\label{fig:histofeaturesMalBen1}  Distributions of the following features computed for malignant and benign masses:
{\it area; perimeter; circularity; mean and standard deviation of normalized radial length;  radial distance entropy; zero crossing; maximum axis}.
}
\end{figure}

\begin{figure}
\begin{center}
\epsfxsize 6.5  true cm
\epsffile{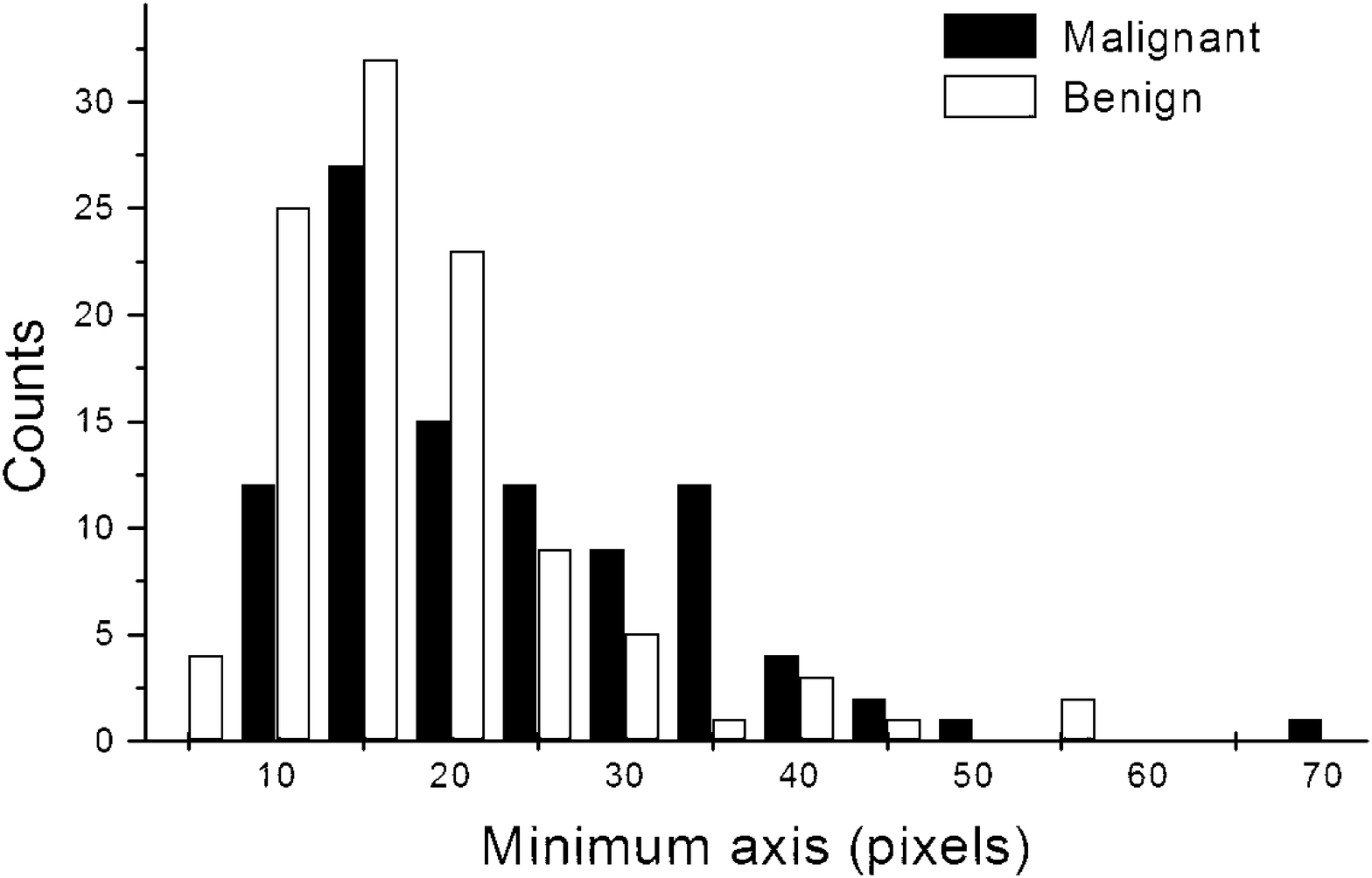}
\epsfxsize 6.5  true cm
\epsffile{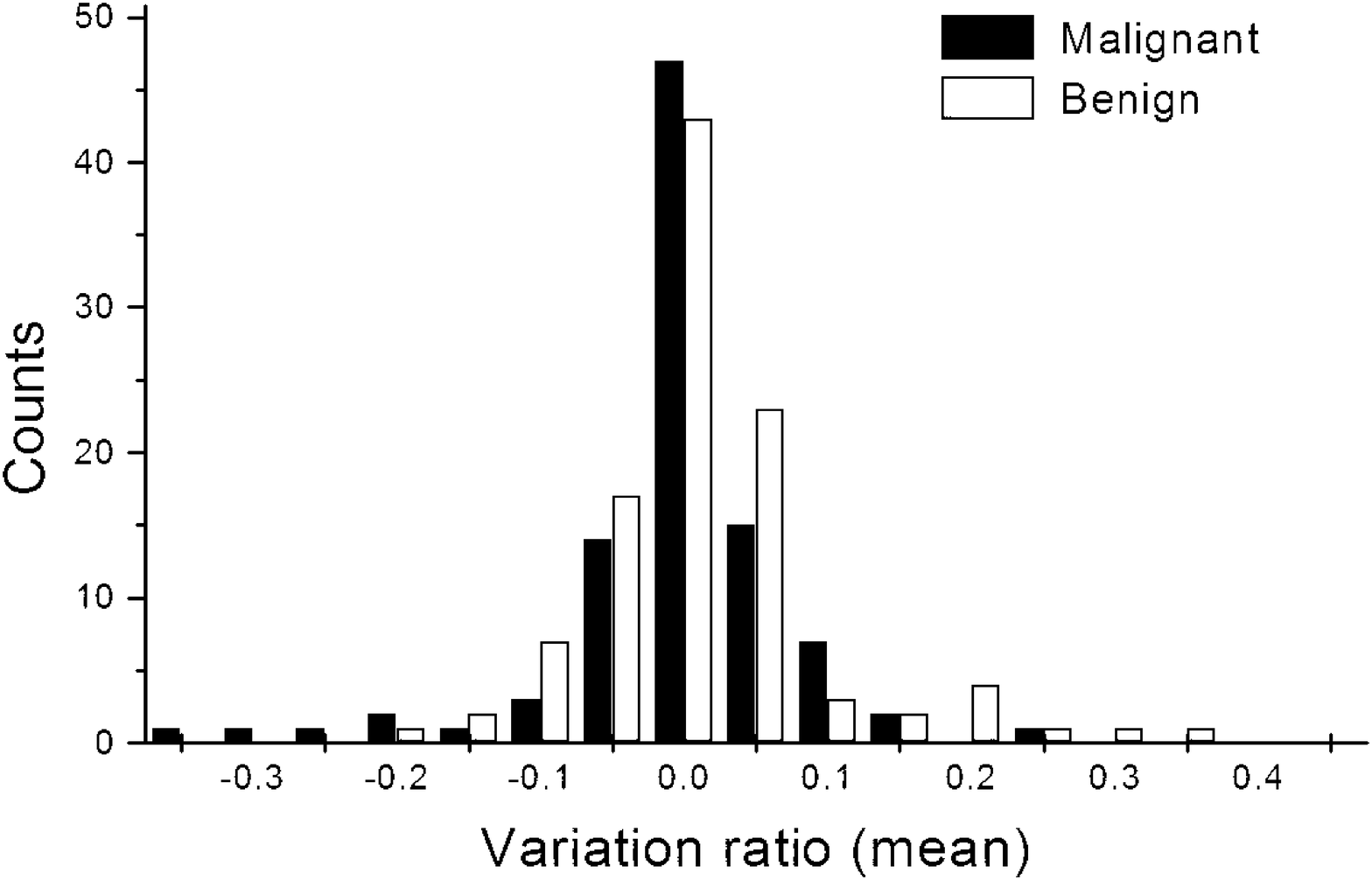}\\
\epsfxsize 6.5  true cm
\epsffile{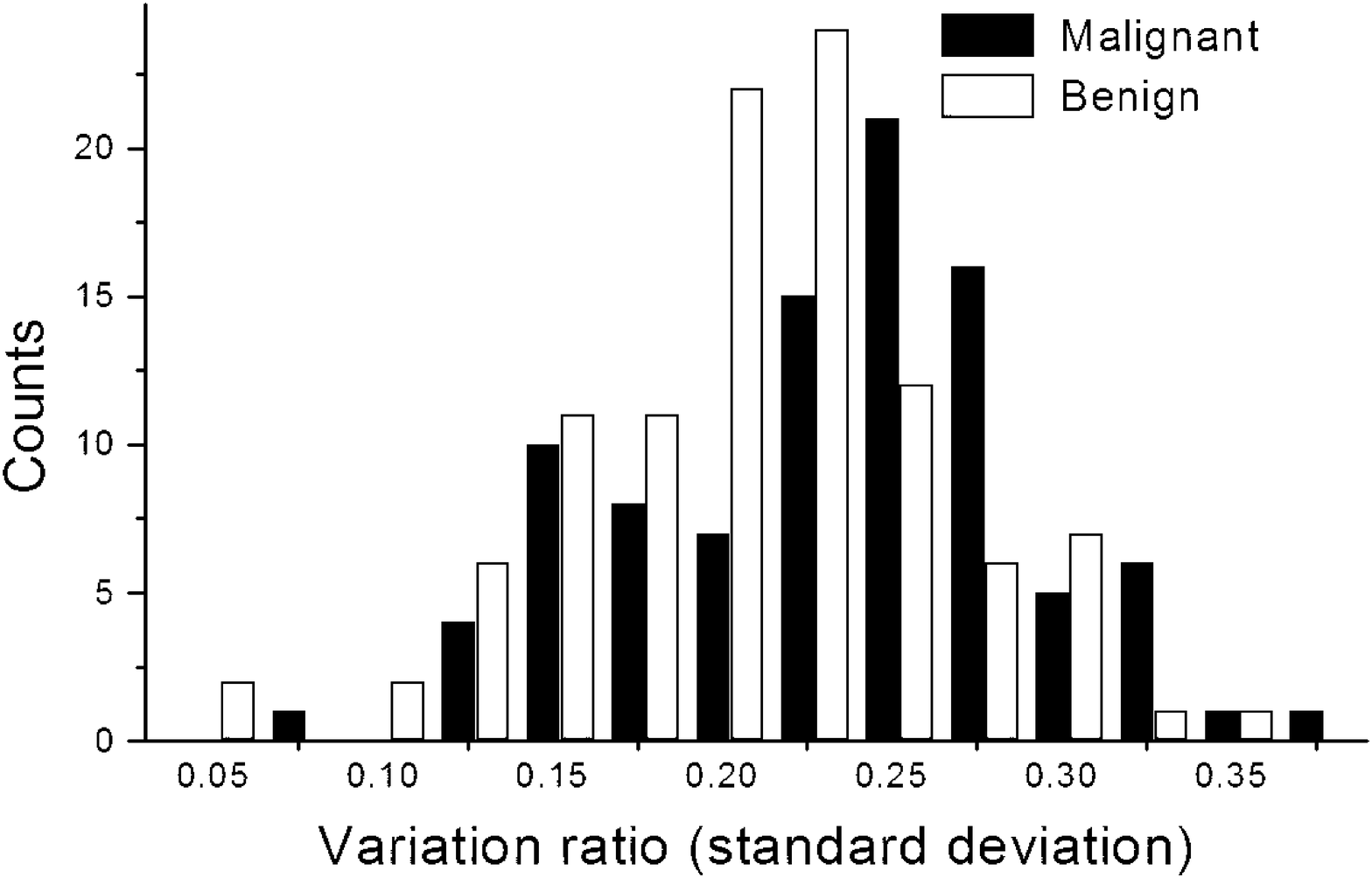}
\epsfxsize 6.5  true cm
\epsffile{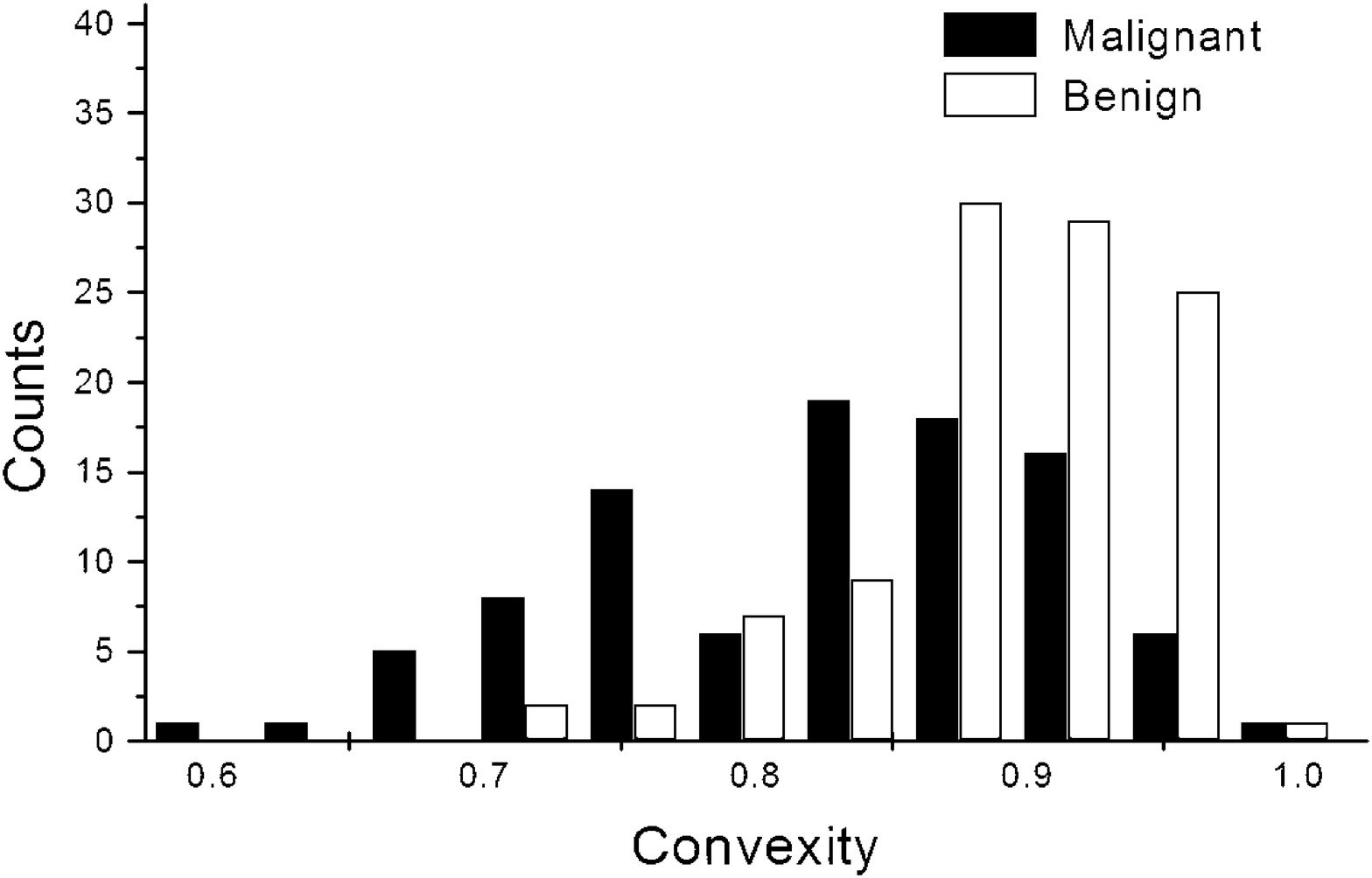}\\
\epsfxsize 6.5  true cm
\epsffile{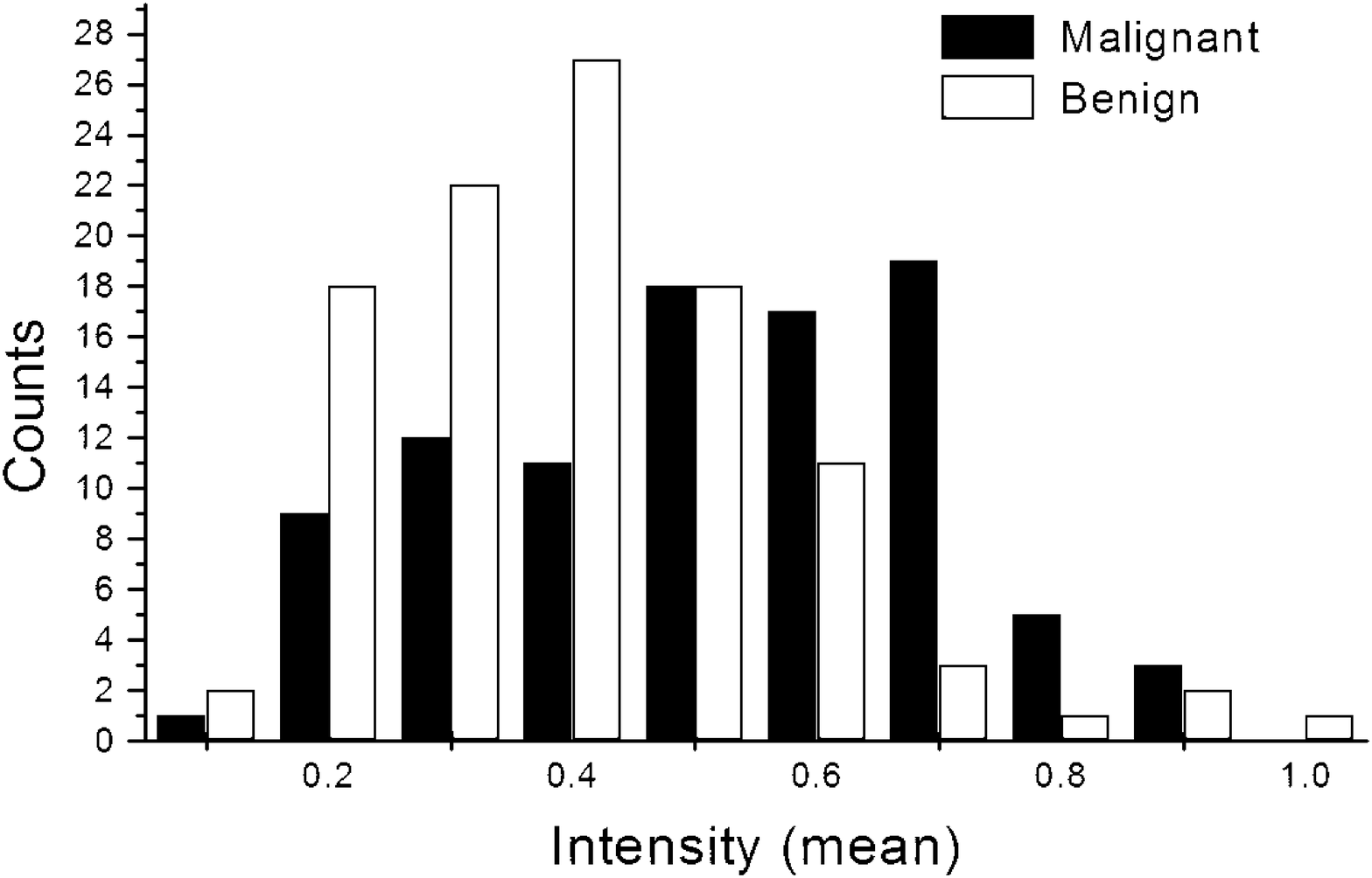}
\epsfxsize 6.5  true cm
\epsffile{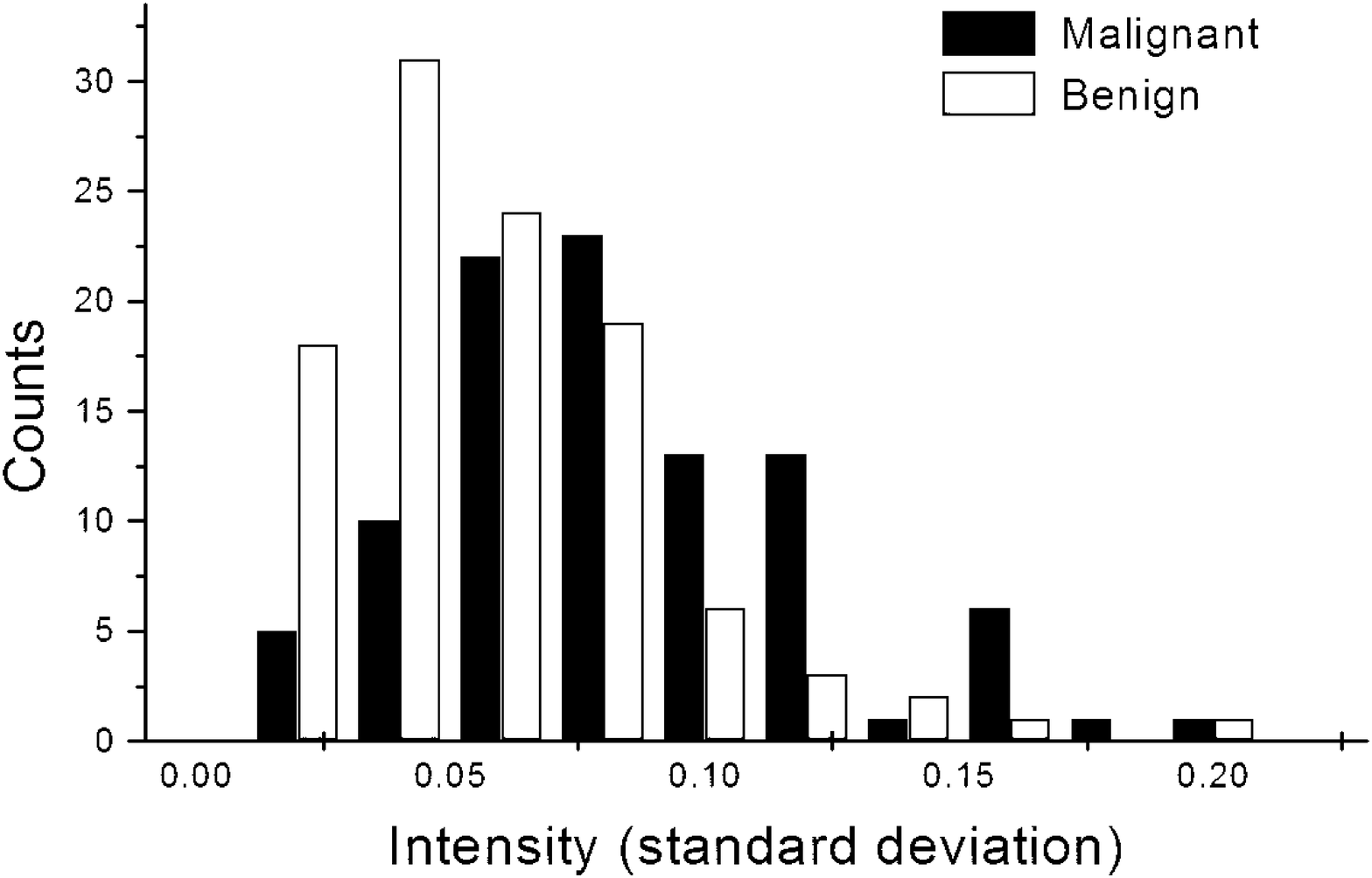}\\
\epsfxsize 6.5  true cm
\epsffile{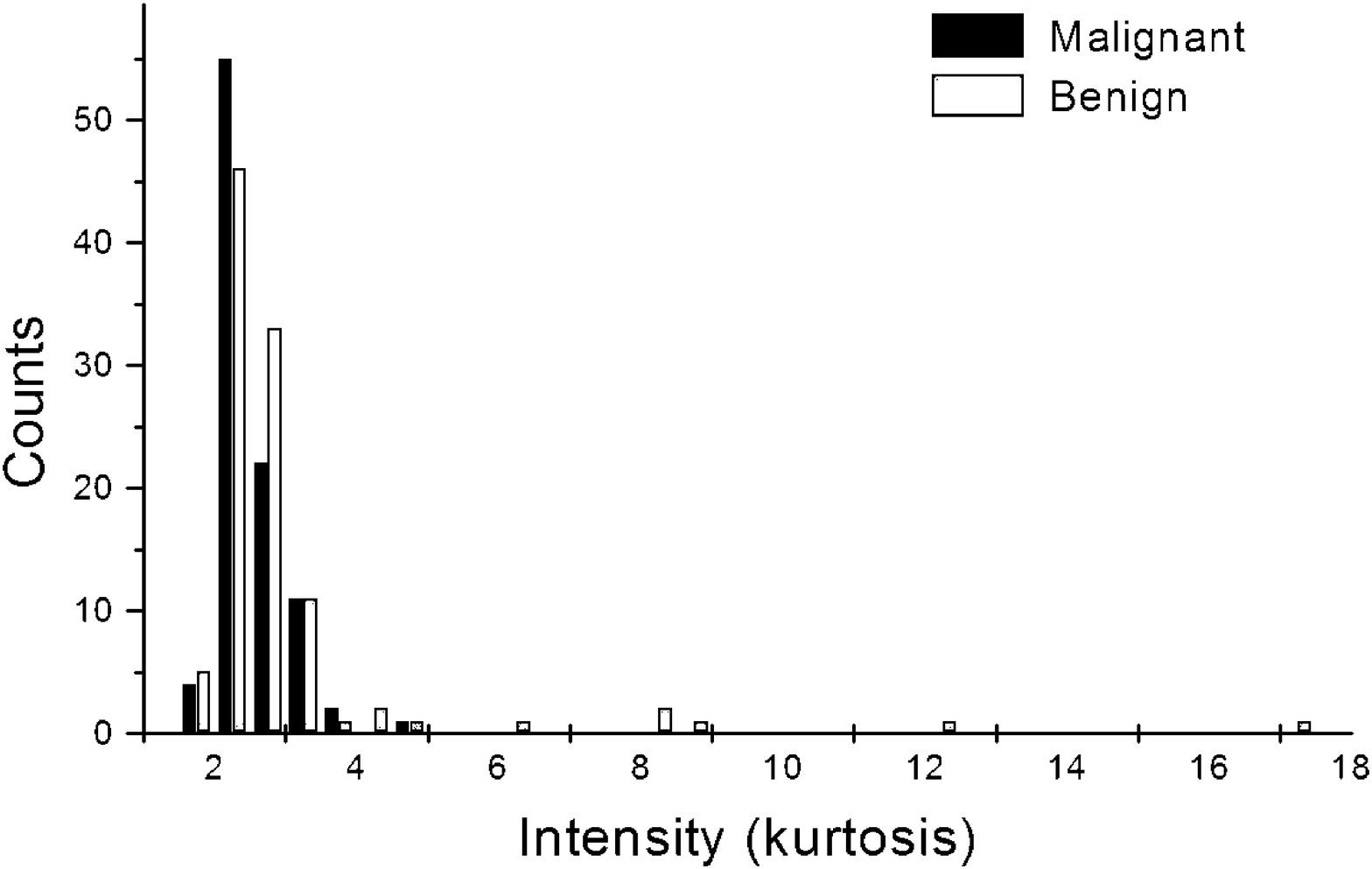}
\epsfxsize 6.5  true cm
\epsffile{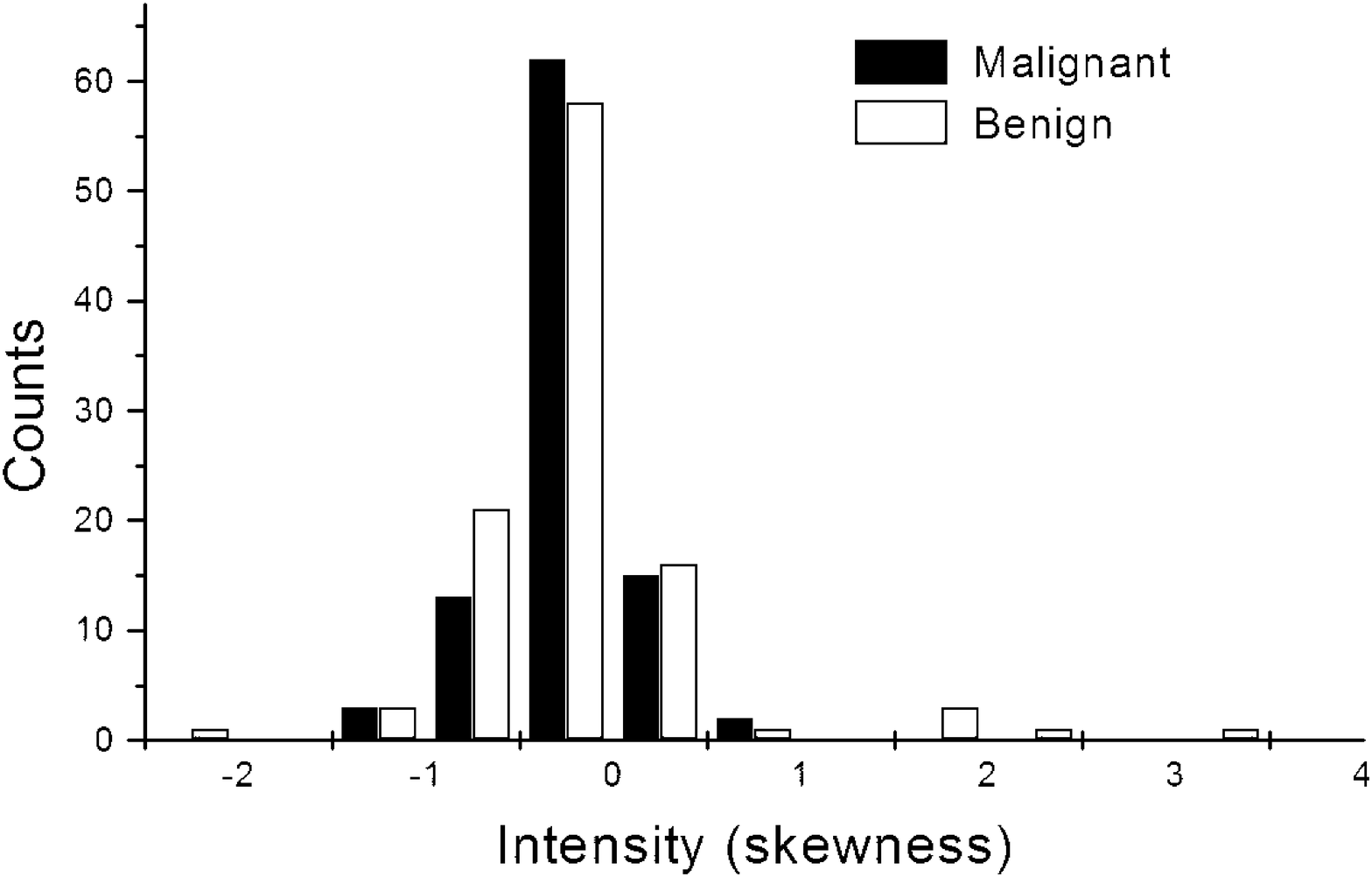}
\end{center}
\caption{\label{fig:histofeaturesMalBen2} Distributions of the following features computed for malignant and benign masses:
{\it minimum axis; mean and standard deviation of variation ratio; convexity; mean, standard deviation, kurtosis and skewness of the mass intensity}.
}
\end{figure}

 The distributions of the features computed for the
malignant and benign cases are plotted  in
figures~\ref{fig:histofeaturesMalBen1} and
\ref{fig:histofeaturesMalBen2}~\footnote{Notice that the shrinking 
factor of 8 applied to both rows and columns of the original image
has not been taken into account in drawing 
the distributions of the quantities 
measured in pixel units.}.  
The mean values of the features extracted from malignant and 
benign masses show highly significant differences 
(p-values$<0.01$)
for 13 features out of the 16 we computed. 
For one of the remaining features (the {\it kurtosis of the intensity}) 
the means are significantly different (p$<$0.05).
The only two features not showing a significant difference in the mean values 
are the {\it mean  of variation ratio} and the {\it skewness of the intensity}.
%
%

The analysis of the linear correlations $\rho(i,j)$ among the 16 features 
(see the matrix of the correlation coefficients in tab.~\ref{tab:correlation}),
lead to the following considerations:
\begin{itemize}
\item the {\it perimeter} is highly correlated to the {\it zero crossing}, the
{\it maximum axis}, the {\it minimum axis} and the {\it area}; if the 
{\it perimeter} is excluded from the feature set, the correlations among
the remaining features all satisfy the constraint 
 $\rho(i,j) < 0.93$; 
\item if we set the threshold   $\rho(i,j) < 0.9$ on the correlation coefficients 
we have to exclude from the set also
the  {\it radial length entropy}, the {\it maximum axis} and the {\it minimum axis} 
ending up with 12 remaining features.
\end{itemize}
\begin{table}[t]
\begin{center}
 \begin{tabular}{|c|cccccccccccccccc|}  \hline
 & 1 & 2 & 3 & 4 & 5 & 6 & 7 & 8 & 9 & 10 & 11 & 12 & 13 & 14 & 15 & 16 \\
\hline
1&100&91&-45&4&-12&-6&88&91&92&0&-8&-19&48&64&4&-25 \\
2&91&100&-71&-14&0&11&98&96&93&0&7&-46&47&63&3&-19 \\
3&-45&-71&100&58&-38&-54&-69&-65&-53&-7&-47&87&-22&-37&-4&-8 \\
4&4&-14&58&100&-68&-84&-8&-16&11&-40&-65&64&10&1&0&-15 \\
5&-12&0&-38&-68&100&91&1&6&-28&-15&78&-50&-17&-11&-5&18 \\
6&-6&11&-54&-84&91&100&10&14&-18&0&82&-62&-13&-7&-2&19 \\
7&88&98&-69&-8&1&10&100&94&92&-10&9&-46&45&61&1&-18  \\
8&91&96&-65&-16&6&14&94&100&89&2&9&-40&49&66&4&-22 \\
9&92&93&-53&11&-28&-18&92&89&100&-1&-16&-25&50&66&4&-24 \\
10&0&0&-7&-40&-15&0&-10&2&-1&100&-20&-1&6&6&7&-6 \\
11&-8&7&-47&-65&78&82&9&9&-16&-20&100&-59&-13&-8&-5&25 \\
12&-19&-46&87&64&-50&-62&-46&-40&-25&-1&-59&100&-10&-26&2&-20 \\
13&48&47&-22&10&-17&-13&45&49&50&6&-13&-10&100&74&-3&-35 \\
14&64&63&-37&1&-11&-7&61&66&66&6&-8&-26&74&100&5&-10 \\
15&4&3&-4&0&-5&-2&1&4&4&7&-5&2&-3&5&100&48 \\
16&-25&-19&-8&-15&18&19&-18&-22&-24&-6&25&-20&-35&-10&48&100 \\
\hline
\end{tabular}
\end{center}
\caption{Matrix of the linear correlation coefficients $\rho(i,j)$ (expressed 
as percentages) among the 16 features. The features are numbered with $i=1,\dots 16$ according to the descriptions given in sec.~\ref{sec:feature_extraction}.
\label{tab:correlation} }
\end{table}

On the basis of these preliminary analysis on the features, we decided to 
select the optimal set of features to be finally computed by the CAD
system by comparing 
the neural network performances obtained with different choices of the
feature set cardinality.

\subsection{Classification}

We prepared 5 different train and test sets for the  5$\times$2
cross validation analysis, by randomly assigning each of the 200
vectors of features   to the train or test set for each of the 5
different trials.

We trained 4 different sets of 10 networks by varying the number of features
taken into account for the classification:
\begin{itemize}
\item[a)] all 16 features are considered;
\item[b)] 15 features are considered (the {\it perimeter}, which has the higher correlations with  other features, is excluded);
\item[c)] 14 features are considered (the  {\it mean value of variation ratio} 
and the {\it skewness of the intensity}  are excluded, as they have the poorest discriminating power);
\item[d)] 12 features are considered (the {\it perimeter},
the  {\it radial length entropy}, the {\it maximum axis} and the {\it minimum axis} 
are excluded, as they have $\rho(i,j)>0.9$ with some other features);
\end{itemize}
The architecture of the three-layer feed-forward neural network we
used consists in $n$ input,
3 hidden and 2 output neurons, $n$ depending on the choice of the
number of features to be classified.
We experimentally observed that the network performances for
all choices of datasets are optimized with 3 neurons in the hidden layer.

Since the classifier
performances and the  comparison among different
classifiers are conveniently evaluated in terms of the area $A_z$
under the ROC curve, 
we reported in tab.~\ref{tab:5X2_Az} the estimated areas under the ROC curves
obtained in each trial.
\begin{table}[t]
\begin{center}
\begin{tabular}{|c|c|c|c|c|c|c|c|}  \hline
& &  {\it a)}  &  {\it b)}  &  {\it c)}  &  {\it d)} &  {\it e)} &  {\it f)} \\
Train & Test & 16 features & 15 features & 14 features & 12 features & 12 features & 12 features \\
set   & set & 200 masses & 200 masses & 200 masses & 200 masses & 221 masses & 226 masses \\
& &  $A_z$ &  $A_z$ & $A_z$ &  $A_z$ &  $A_z$ &  $A_z$ \\
\hline
$A_1$ & $B_1$ & 0.839 & 0.845 & 0.801 & 0.849 & 0.827 & 0.809 \\
$B_1$ & $A_1$ & 0.815 & 0.789 & 0.794 & 0.829 & 0.810 & 0.804 \\
$A_2$ & $B_2$ & 0.807 & 0.805 & 0.796 & 0.803 & 0.788 & 0.777 \\
$B_2$ & $A_2$ & 0.811 & 0.789 & 0.796 & 0.802 & 0.786 & 0.779 \\
$A_3$ & $B_3$ & 0.799 & 0.831 & 0.776 & 0.830 & 0.795 & 0.790 \\
$B_3$ & $A_3$ & 0.769 & 0.770 & 0.756 & 0.760 & 0.745 & 0.737 \\
$A_4$ & $B_4$ & 0.791 & 0.828 & 0.788 & 0.810 & 0.797 & 0.792 \\
$B_4$ & $A_4$ & 0.831 & 0.804 & 0.810 & 0.795 & 0.785 & 0.780 \\
$A_5$ & $B_5$ & 0.806 & 0.819 & 0.786 & 0.816 & 0.786 & 0.782 \\
$B_5$ & $A_5$ & 0.765 & 0.765 & 0.766 & 0.756 & 0.753 & 0.748 \\
 \hline           
\multicolumn{2}{|c|}{$A_z$ mean}     & 0.803 & 0.805 & 0.787 & 0.805 & 0.787 & 0.780 \\
\multicolumn{2}{|c|}{$\sigma_{A_z}$} & 0.024 & 0.026 & 0.017 & 0.030 & 0.024 & 0.023 \\
\multicolumn{2}{|c|}{$A_z$ min}  & 0.765 & 0.765 & 0.756 & 0.756 & 0.745 & 0.737 \\
\multicolumn{2}{|c|}{$A_z$ max}  & 0.839 & 0.845 & 0.810 & 0.849 & 0.827 & 0.809 \\
\hline
\end{tabular}
\end{center}
\caption{Evaluation of the performances of the 
neural classifiers trained on  feature sets with different 
cardinalities: the  $A_z$ values obtained on the test sets
according to the  5$\times$2 cross validation method are reported.
\label{tab:5X2_Az}}
\end{table}
The average $A_z$ obtained on each test set and the standard 
deviation referred to the 10 different trials for each set of 
features are reported.
As can be noticed, the performances the neural classifiers achieve are
robust, i.e. almost independent of the partitioning of the available
data into the train and test sets.  
The mean $A_z$ value obtained in classifying 16, 15 and 12 features, 
i.e. in the {\it a)},  {\it b)} and  {\it d)} cases, are indeed very similar (p-values$>$0.05).
Even in the    {\it c)} case, where the less discriminating 
features are excluded from the set of features, the
mean $A_z$ value is not significantly different from those obtained 
with different choices of the features to be taken into account (p-values$>$0.05).

We can conclude that, being smaller but as predictive as the other sets 
of features, the   {\it d)} set constituted by the 12 less correlated features  
will be considered as the optimal set of features to be 
extracted from segmented masses to determine their likelihood of 
malignancy.  
  
\begin{figure}
\begin{center}
\epsfxsize 10.  true cm
\epsffile{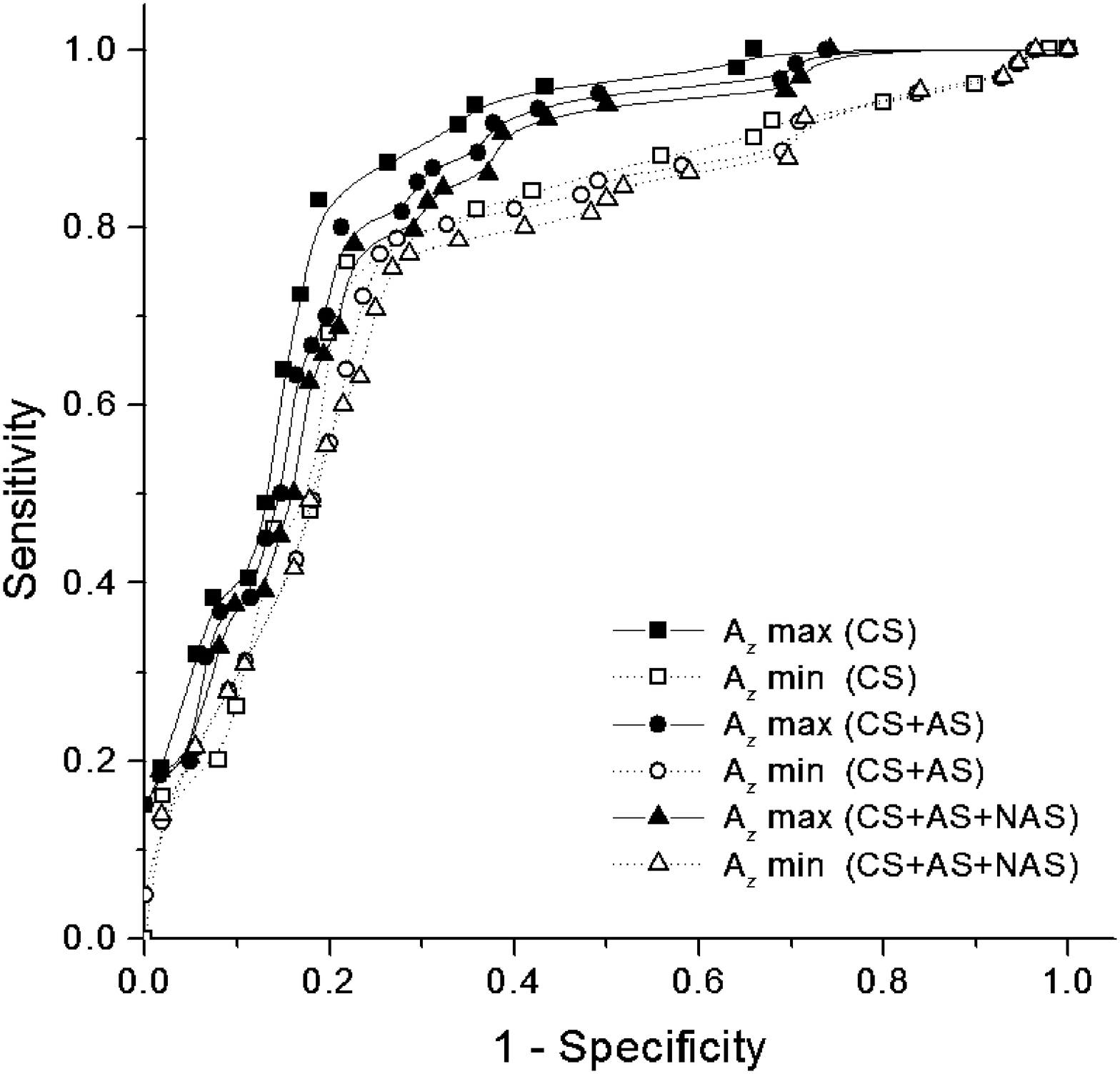}
\end{center}
\caption{\label{fig:ROC} ROC curves obtained in the classification
of 12 features (see tab.~\ref{tab:5X2_Az}) extracted from the datasets 
of: {\it correctly-segmented} 
masses (CS);  {\it correctly-segmented} and {\it acceptably-segmented}  masses (CS+AS);
 {\it correctly-segmented}, {\it acceptably-segmented} and {\it non-acceptably-segmented}  masses  (CS+AS+NAS).}
\end{figure}

The mean $A_z$ value obtained with this system configuration on the sets of
{\it correctly-segmented} masses is 
$A_z=0.805\pm 0.030$.
The ROC curves realizing the minimum and maximum $A_z$
values (0.756 and 0.849 respectively) are reported in fig.~\ref{fig:ROC},
where they are compared to the corresponding ROC curves 
achievable by running the CAD on the dataset of the 
{\it correctly-segmented} masses added by the 21 
{\it acceptably-segmented} ones (column {\it e)} in  tab.~\ref{tab:5X2_Az}) and finally 
including also the 5 {\it non-acceptably-segmented} cases (column {\it f)} in  tab.~\ref{tab:5X2_Az}).

As shown  in fig.~\ref{fig:ROC}, in the   {\it d)} case 
high values as 80--85\% of sensitivity to malignant masses 
correspond to specificity values in the 70--80\% range.

\subsection{The reject option}

As the segmentation algorithm leads to three classes of segmented masses according to the radiologist's evaluation, 
we considered the opportunity of 
exploiting this quality control performed on the segmentation 
step of the analysis 
to  minimize the amount of cases misclassified  by the  
neural decision-making system.
We adopted the {\it reject option}~\cite{Vento,Vasconcelos}, i.e. we evaluated the convenience of not assigning a class to the input sample (rejection of the sample), rather than risking a wrong classification. 
A suitable criterion for rejection has to reject the highest possible percentage of samples which would be otherwise misclassified.
The {\it reject option} is based on an estimate of the classification reliability, measured by a {\it reliability evaluator} $\Psi$.
Once a {\it reject threshold} $\sigma$ has been fixed, a sample is rejected if the corresponding value of $\Psi$ is below $\sigma$.
We set a trivial correspondence between the values assumed by the 
$\Psi$ function and the
radiologist's opinion about the quality of mass segmentation:
$\Psi=1$ for {\it correctly-segmented} masses, $\Psi=0.5$ for {\it acceptably-segmented} masses and $\Psi=0$ for {\it non-correctly-segmented} masses. 
In other words,  the $\Psi$ function 
is  directly determined  by the radiologist,  rather than being implemented as a function to be automatically derived from data.
Only two possible classes of  values for the threshold $\sigma$  make sense in this case: $0<\sigma_{\rm a}<0.5$ and  $0.5<\sigma_{\rm b}<1$.  
If $\sigma_{\rm a}$ is set as {\it reject threshold}, the {\it correctly-segmented} masses and the {\it acceptably-segmented} ones (corresponding to the $\epsilon_{\rm CS+AS}$ fraction of the dataset) will be classified by the system; otherwise, if  $\sigma_{\rm b}$ is chosen, only the {\it correctly-segmented} masses
(the   $\epsilon_{\rm CS}$ fraction of the cases) will be classified, whereas the remaining cases will be rejected.

As reported in columns {\it d)} and {\it e)}  of tab.~\ref{tab:5X2_Az}, the mean $A_z$ value obtained in case the $\epsilon_{\rm CS+AS}$ fraction of cases is classified does not significantly differ from the mean $A_z$ computed on the $\epsilon_{\rm CS}$ fraction of cases (p=0.16). It means that it is not worth setting a {\it reject threshold} as severe as  $\sigma_{\rm b}$, since it will only lead to a reduction of the rate of reliably classified cases, without improving the classification reliability.
%
By contrast, the comparison between the mean $A_z$ value obtained on the sets including the  {\it non-correctly-segmented} masses reported in column {\it f)} and the mean $A_z$ of column {\it d)} shows a statistically significant difference (p=0.048).
Despite this conclusion is achieved on the basis of a borderline p value, the choice of   $\sigma_{\rm a}$ as {\it reject threshold} will preserve the system performances from a slight decrease due to the possible failure of 
the mass segmentation step. 
Once the  {\it reject threshold}  $\sigma_{\rm a}$ is applied to the system, 
a small subset of masses will be initially rejected as the segmentation algorithm did not lead to an
acceptable result; however, the radiologist in this case has the advantage that he
can more safely trust the stability and reliability of the CAD performances on the 
remaining $\epsilon_{\rm CS+AS}$ fraction of the cases.


\section{Conclusions and discussion}

We developed a CAD system for the classification of mammographic
masses into malignant and benign with the aim of supporting
radiologists in the visual diagnosis of the degree of
mass malignancy. 
Several expert systems with a similar purpose have been recently discussed in
the literature. 
In the paper by Timp  and  Karssemeijer~\cite{Timp}
the influence of the segmentation method on the performance of a CAD system was investigated, obtaining 
$A_z=0.74,0.72$ and $0.67$ for segmentation based on dynamic programming, on the discrete contour model and on region growing, respectively.
Kinnard {\it et al.}~\cite{Wang} studied the efficacy of image features versus likelihood features of tumor boundaries for differentiating benign and malignant tumors; a region growing technique was implemented in the mass segmentation and two different neural networks were adopted in the classification. Different combinations of these components led to  
$A_{z}=0.66,0.71$ and $0.84$,  respectively.
Mudigonda {\it et al.}~\cite{Mudigonda} compared the discriminating capabilities of gradient-based and texture-based features. Different classification approaches on different sets of mammograms led to $A_{z}=0.85,0.67,0.6,0.76,0.52$ and $0.73$.
Hadjiski {\it et al.}~\cite{Hadjiski2} developed a hybrid classifier by combining an 
unsupervised model based on an adaptive resonance theory network and a
supervised linear discriminant classifier. They reached the value $A_{z}=0.81$
for the hybrid classifier to be compared to $A_{z}=0.78$ and 0.80 obtained
by means of the linear discriminant classifier alone and a back-propagation neural network respectively.
Huo {\it et al.}~\cite{Huo} implemented a region growing technique to segment masses and extracted radial edge-gradient information to gather the mass malignancy, obtaining $A_z=0.85$.
Sahiner {\it et al.}~\cite{Sahiner2} developed a three-stage segmentation method based on  clustering, active contour and spiculation detection.
They evaluated the improvement the extraction of morphological features can lead to a mass classification based on texture features extracted from a band of pixels surrounding the mass. They obtained 
$A_z=0.83\pm 0.02, 0.84\pm 0.02$ and $0.87\pm 0.02$ on morphological, texture and combined features, respectively.
They also combined the analysis of different views of a mass, obtaining
$A_z=0.91\pm 0.02$.
 The issue of improving the classification performances in the mass diagnosis is discussed in the paper by
Lim and Er~\cite{Lim}, where  generalized dynamic fuzzy neural networks are introduced.
In this case, the most appropriate structure for the classifier is automatically obtained by means of a self-adapting of 
the network structure during the learning process.
 In classifying mammographic masses they obtain  
$A_z=0.868\pm 0.020$.
Hadjiiski {\it et al.}~\cite{Hadjiiski} exploited the interval change information to evaluate the mass malignancy.
The information on the prior image significantly  improved the accuracy of mass classification from $A_{z}=0.82$ to
$A_{z}=0.88$.
%
Another paper by Sahiner {\it et al.}~\cite{Sahiner} discussed the effect of mass segmentation on characterization:
texture, morphological and spiculation features were extracted from masses segmented by a computerized technique and by the radiologist, obtaining $A_{z}=0.89$ and $0.88$, respectively.
Sahiner   {\it et al.} in a different work~\cite{Sahiner0} transformed 
a band of pixels surrounding a segmented mass into the Cartesian plane. They computed and classified texture features of the transformed images, ending up with  $A_z=0.94$.
In another study by Huo {\it et al.}~\cite{Huo1},
three different automated classifiers were used to 
merge various  features related to the margin and density of the masses 
into a  likelihood of malignancy, obtaining $A_z=0.94$.

Despite the area  $A_z$ under the ROC curve provides a 
good measuring instrument to make a comparison among the performances 
of different CAD systems, 
the reliability of this measurement depends also on the dataset 
used to train and test the CAD.
It is very  difficult in most cases to evaluate how a particular 
database is well populated, i.e. whether it is sufficiently
representative of each possible appearance of the 
pathology one aims to detect.
Even the partitioning of the database into train and test sets 
is not a trivial task: one should at least {\it a posteriori} verify
the robustness of the results achieved with
respect to different data partitioning.

The CAD system we present in this paper has been developed according to 
all these considerations. The performances it reaches are in the same 
range of most of the previously reported results. 
We devoted the main efforts to the development of the mass segmentation step.
We significantly improved the procedure proposed by Chen {\it et al.}~\cite{Chen} for the analysis of sonograms, 
by  making  the algorithm iterative. In particular, as mammographic images 
show a better resolution with respect to sonograms, an appropriate segmentation procedure 
needs to be particularly sensitive even to subtle variations occurring in the mass margins. 
The iterative procedure we propose is able to identify even very small arms and branches 
possibly occurring especially in malignant masses.
Furthermore, it has the ability
of handling masses with various sizes: there is no size limit even
up to few pixels. The algorithm execution  is not
computationally expensive; computation time has linear relation
with the size of the mass. 
In contrast with the analysis by Chen {\it et al.}~\cite{Chen} we could not find any improvement in the 
mass margin definition and consequently in the final CAD results by 
introducing the wavelet transforms in the identification of the 
critical points defining the mass margin. 

Despite the accuracy of segmentation algorithms is usually evaluated in
terms of the overlap between the area segmented by the CAD and the manual segmentation
of the mass provided by an experienced radiologist, we have not had the possibility 
to carry out such a study.
To test the reliability of our segmentation procedure we adopted  
the following criterion:
the  radiologist partitioned the segmented masses in three classes, according to 
a decreasing rate of reliability of the segmentation result;
the CAD performances were evaluated first on the dataset of 
{\it correctly-segmented} masses, obtaining $A_z=0.805\pm 0.030$;
then, also {\it acceptably-segmented} and {\it non-correctly-segmented}  masses 
(which represent however small fractions of the dataset) were added in two 
steps to the test sets, obtaining $A_z=0.787\pm 0.024$ and $0.780\pm 0.023$, 
respectively. 
As the difference among the $A_z$ values obtained on 
the fraction $\epsilon_{\rm CS}=88.5\%$ of {\it correctly-segmented} masses and the fraction  $\epsilon_{\rm CS+AS}=97.8\%$ of {\it correctly-} and {\it acceptably-segmented} cases is not statistically significant, 
we can conclude that even in case the segmentation does 
not provide an extremely refined identification of the mass margin, the interplay
between the morphological and textural features
extracted from the segmented area still leads to a 
reliable classification result on the 97.8\% of the database.

\section*{Acknowledgments}

We would like to thank the professors, the radiologists and the employees of the Radiological Departments who contributed to the acquisition of the mammographic database in the framework of the INFN-founded CALMA Project.
We are grateful to Dr S. Franz from ICTP (Trieste, Italy) for useful suggestions and discussions.  Special thanks to Dr M.  Tonutti from Cattinara Hospital (Trieste, Italy) for her essential contribution in the evaluation of the efficiency of the segmentation algorithm.


\begin{thebibliography}{99}
%
%
\bibitem{Greenlee} R.T. Greenlee, M.B. Hill-Harmon, T. Murray and M. Thun, Cancer statistics, {\it Ca-Cancer J Clin} {\bf 51}(1) 15--36 (2001). Erratum in: {\it Ca-Cancer J Clin} {\bf 51}(2) 144 (2001).
\bibitem{Levi}
F. Levi, F. Lucchini, E. Negri, W. Zatonski, P. Boyle and C. La Vecchia,
Trends in cancer mortality in the European Union and accession
countries, 1980-2000, {\it Ann. Onc.}  {\bf  15}(9) 1425--1431 (2004).
\bibitem{Smith} R.A. Smith,  The epidemiology of breast cancer, { \it Syllabus: A Categorical Course in Breast Imaging},  Radiological Society of North America, pp.  7--20. D.B. Kopans and E.B. Mendelson eds (1995).
\bibitem{Landis} S.H. Landis, T. Murray, S. Bolden and P.A. Wingo, Cancer statistics, 1999. {\it Ca-Cancer J Clin} {\bf 49}(1) 8--31 (1999).
\bibitem{Zuckerman} H.C. Zuckerman, The role of mammography in the diagnosis of breast cancer, {\it   Breast cancer: diagnosis and treatment}, pp. 152--72.  I.M. Ariel and J.B. Cleary eds, New York: McGraw-Hill (1987).
\bibitem{Haus} A.G. Haus and M.J. Yaffe  eds, {\it Syllabus: A Categorical Course in Physics - Technical Aspects in Breast Imaging}, Radiological Society of North America, Presented at the 79$^{\rm th}$ Scientific Assembly and Annual Meeting of RSNA (1993).
\bibitem{Adler} D.D. Adler and M.A. Helvie, Mammographic biopsy recommendations, {\it Curr Opin Radiol} {\bf 4}(5) 123--9 (1992).
\bibitem{Huo1} Z. Huo, M.L. Giger, C.J. Vyborny, D.E. Wolverton, R.A. Schmidt and K. Doi,  Automated computerized classification of malignant and benign masses on digitized mammograms,
{\it Acad Radiol} {\bf 5}(3) 155--68 (1998).
\bibitem{Huo2} Z. Huo,  M.L. Giger, C.J. Vyborny, D.E. Wolverton and C.E. Metz,
 Computerized classification of benign and malignant masses on digitized mammograms: a study of robustness,
{\it Acad Radiol} {\bf 7}(12) 1077--84 (2000).
\bibitem{Sahiner0}
 B. Sahiner, H.P. Chan, N. Petrick, M.A. Helvie and M.M. Goodsitt,
 Computerized characterization of masses on mammograms: the rubber band straightening transform and texture analysis,
{\it Med Phys} {\bf 25}(4) 516--26 (1998).
\bibitem{Wirth} M.A. Wirth and A. Stapinski,  Segmentation of the
Breast Region in Mammograms Using Active Contours, {\it Visual Communications and Image Processing} {\bf 5150}, pp. 1995--2006. T. Ebrahimi and T. Sikora eds, Lugano, Switzerland (2003).
\bibitem{Amini} A. Amini, S. Tehrani and T. Weymouth, Using Dynamic
programming for Minimizing the Energy of Active Contours in the
Presence of Hard Constraints, {\it Proc Second Int Conf Computer Vision}, pp.  95--99. Tarpon Springs,  FL (1988).
\bibitem{Sahiner} B. Sahiner, N. Petrick, H.P. Chan,
L.M.  Hadjiiski, C. Paramagul, M.A. Helvie and M.N. Gurcan,
 Computer-Aided Characterization of Mammographic Masses:
Accuracy of Mass Segmentation and Its Effects on
Characterization, {\it IEEE Trans Med Imaging} {\bf 20}(12) 1275--84 (2001).
\bibitem{Matthew} M.A. Kupinski and M.L. Giger,
Automated Seeded Lesion Segmentation on Digital
Mammograms, {\it IEEE Trans Med Imaging} {\bf  17}(4) 510--17 (1998).
\bibitem{Timp}
S.~Timp and N.~Karssemeijer, A new 2D segmentation method based on dynamic   programming applied to computer aided detection in mammography, {\it Med Phys} {\bf 31}(5) 958--71 (2004).
\bibitem{Chen} D.R. Chen, R.F. Chang, W.J. Kuo, M.C. Chen and
Y.L. Huang, Diagnosis of Breast Tumors with Sonographic Texture
Analysis using Wavelet Transform and Neural Networks, {\it
Ultrasound Med Biol} {\bf 28}(10) 1301--10 (2002).
\bibitem{Wang} L. Kinnard, S.B. Lo, P. Wang, M.T. Freedman, M. Chouikha, Separation of Malignant and Benign Masses Using Image and Segmentation Features, {\it Medical Imaging 2003: Image Processing} {\bf 5032} 835--842 (2003).
\bibitem{Kinnard} L.M. Kinnard, S.C.B. Lo, P.C. Wang, M.T. Freedman and M.F. Chouikha, Separation of malignant and benign masses using maximum-likelihood modeling and neural networks, {\it Proc. SPIE} {\bf 4684}  pp. 733--741. Medical Imaging 2002: Image Processing, M. Sonka and J.M. Fitzpatrick eds (2002).
\bibitem{Christoyianni} I. Christoyianni, E. Dermatas and G. Kokkinakis, Neural Classification of Abnormal Tissue in Digital Mammography using Statistical Features of the Texture, {\it IEEE, ICECS'99} {\bf 1} 117--120 (1999).
\bibitem{Qian} W. Qian, L. Li and L.P. Clarke,
Image Feature Extraction for Mass Detection in Digital Mammography: Influence of Wavelet Analysis, {\it Med Phys} {\bf 26} 402--408 (1999).
\bibitem{Hadjiiski}
L. Hadjiiski, B. Sahiner, H.P. Chan, N. Petrick, M.A. Helvie and M. Gurcan,
 Analysis of temporal changes of mammographic features: computer-aided classification of malignant and benign breast masses,
{\it Med Phys} {\bf 28}(11) 2309--17 (2001).
\bibitem{Sahiner2}
B. Sahiner, H.P. Chan, N. Petrick, M.A. Helvie and L.M. Hadjiiski,
 Improvement of mammographic mass characterization using spiculation measures and morphological features,
{\it Med Phys} {\bf 28}(7) 1455--65 (2001).
\bibitem{Mudigonda}
N.R.~Mudigonda, R.M.~Rangayyan and J.E.~Desautels, Gradient and texture analysis for the classification of mammographic masses,
{\it IEEE Trans Med Imaging} {\bf 19}(10) 1032--43 (2000).
\bibitem{Huo}
Z. Huo, M.L. Giger, C.J. Vyborny, U. Bick, P. Lu, D.E. Wolverton and R.A. Schmidt, Analysis of spiculation in the computerised classification of mammographic masses, {\it Med Phys} {\bf 22}(10) 1569--1579 (1995).
\bibitem{Birads}
American College of Radiology. Breast Imaging - Reporting and Data
System (BI-RADS\textregistered), 3$^{\rm rd}$ ed. Reston, VA: American College of Radiology (1998), 
http://www.birads.at/
%
\bibitem{Lambrou} T. Lambrou, A.D. Linney, R.D. Speller, A. Todd-Pokropek, Statistical Classification of Digital Mammograms Using Features from the Spatial and Wavelet Domains,
{\it 2$^{\rm nd}$ European Medical and Biological Engineering Conference
(EMBEC)}, Vienna, Austria (2002).
\bibitem{Bruce} L.M. Bruce and R.R. Adhami, Classifying Mammographic Mass Shapes Using the Wavelet Transform Modulus-Maxima Method, {\it IEEE Trans Med Imaging}, {\bf 18}(12) 1170--77 (1999).
\bibitem{Wei} D. Wei, H. Chan, M.A. Helvie, B. Shiner, N. Petrick, D.D. Adler and M.M. Goodsitt, Classification of Mass and Normal Breast Tissue on Digital Mammograms: Multiresolution Texture analysis,
{\it Med Phys} {\bf 22}(9) 1501--13 (1995).
\bibitem{Sheng} S. Liu, C.F. Babbs and E.J. Delp, Multiresolution Detection of Spiculated Lesions in Digital
Mammograms, {\it IEEE Trans Image Processing} {\bf 10}(6) 874--884 (2001).
\bibitem{Arbach} L. Arbach, D.L. Bennett, J.M. Reinhardt and G. Fallouhc , Classification of Mammographic Masses: Comparison between Backpropagation Neural Network (BNN) and Human Readers,
{\it Medical Imaging} {\bf 5032} 810--818 (2003).
\bibitem{Lauria} A. Lauria, R. Palmiero, P. Cerello, B. Golosio, F. Fauci, R. Magro, G. Raso, S. Tangaro, P.L. Indovina, The CALMA system: an Artificial Neural Network Method for Detecting Masses and Microcalcifications in Digitized Mammograms,  {\it Nucl Instr Meth A} {\bf 518}  391--393 (2004).
\bibitem{Bovis} K. Bovis and S. Singh, Classification of Mammographic Breast Density Using a Combined Classifier Paradigm,
{\it Medical Image Understanding and Analysis (MIUA) Conference},
Portsmouth (2002).
\bibitem{Hadjiski2}
L.~Hadjiski, B.~Sahiner, H.P.~Chan, N.~Petrik  and M.~Helvie, Classification of Malignant and Benign Masses Based on Hybrid ART2LDA Approach, {\it IEEE Trans Med Imaging} {\bf 18}(12) 1178--87 (1999).
\bibitem{Lim} 
W.K.~Lim and M.J.~Er,  Classification of mammographic masses using generalized dynamic fuzzy neural networks,
{\it Med Phys} {\bf 31}(5) 1288--95 (2004).
%
\bibitem{Haykin}  S. Haykin, {\it Neural Networks: a Comprehensive Foundation}, Macmillan College Publishing Company, New York, 1994.
\bibitem{Dietterich} T.G. Dietterich, Approximate Statistical Test For Comparing Supervised Classification Learning Algorithms, {\it Neural Computation} {\bf 10}(7) 1895--923 (1998).
\bibitem{Stone} M. Stone, Cross-validatory choice and assessment of statistical predictions, {\it  Journal of Royal Statistical Society B} {\bf  36} 111--147 (1974).
\bibitem{Metz} C.E. Metz,  ROC methodology in radiologic imaging, {\it Invest Radiol}  {\bf 21}(9) 720--33 (1986).

%
\bibitem{Bottigli} U. Bottigli, P. Delogu, M.E. Fantacci, F. Fauci, B. Golosio, A. Lauria, R. Palmiero, G. Raso, S. Stumbo and S. Tangaro, Search of microcalcification clusters with the CALMA CAD station, {\it The International Society for Optical Engineering (SPIE)}  {\bf 4684} 1301--10 (2002).
\bibitem{magic5} R. Bellotti, S. Bagnasco, U. Bottigli, M. Castellano, R. Cataldo, E. Catanzariti, P. Cerello, S.C. Cheran,
F. De Carlo, P. Delogu, I. De Mitri, G. De Nunzio, M.E. Fantacci, F.
Fauci, G. Forni, G. Gargano, B. Golosio, P.L. Indovina, A. Lauria,
E. Lopez Torres, R. Magro, D. Martello, G.L. Masala, R. Massafra, P.
Oliva, R. Palmiero, A. Preite Martinez, R. Prevete, M. Quarta, L.
Ramello, G. Raso, A. Retico, M. Santoro, M. Sitta, S. Stumbo, S.
Tangaro, A. Tata and E. Zanon, The MAGIC-5 Project: Medical
Applications on a Grid Infrastructure Connection, {\it  IEEE
Nuclear Science Symposium Conference Record} Vol. {\bf 3} 1902--1906 (2004).
%
%
%
\bibitem{Vento}
C.~De Stefano, C.~Sansone and M.~Vento, To reject or not to reject: that is the question-an answer in case of neural classifiers, {\it IEEE Transactions on
Systems, Man and Cybernetics, Part C}  
{\bf 30}(1) 84--94 (2000).
\bibitem{Vasconcelos}
G.C.~Vasconcelos, M.C.~Fairhust and D.L.~Bisset, Investigating feedforward neural networks with respect to the rejection of spurious patterns, {\it Pattern Recognit Lett} {\bf 16}(2) 207--212 (1995).
%
\end{thebibliography}
\end{document}